\documentclass[12pt]{article}
\usepackage{epsfig}
\usepackage{amssymb}

\def\a{\alpha}
\def\b{\beta}

\def\g{\gamma}
\def\h{\eta}

\def\k{\kappa}

\def\m{\mu}
\def\n{\nu}

\def\p{\pi}

\def\r{\rho}
\def\s{\sigma}
\def\t{\tau}

\def\D{\Delta}

\def\G{\Gamma}

\def\O{\Omega}

\def\ve{\varepsilon}

\def\dg{\dagger}                                     % hermitian conjugate
\def\mt{\widetilde{m}_1}
\def\mb{\overline{m}}
\def\VEV#1{\left\langle #1\right\rangle}        % < >
\def\beq{\begin{equation}}
\def\eeq{\end{equation}}

\def\bea{\begin{eqnarray}}
\def\eea{\end{eqnarray}}
\def\NO{\nonumber}
\def\Bar#1{\overline{#1}}

\def\pl#1#2#3{Phys.~Lett.~{\bf B {#1}} ({#2}) #3}
\def\np#1#2#3{Nucl.~Phys.~{\bf B {#1}} ({#2}) #3}

\def\pr#1#2#3{Phys.~Rev.~{\bf D {#1}} ({#2}) #3}

\begin{document}
\date{}

\title{
{\normalsize
\mbox{ }\hfill
\begin{minipage}{3cm}
DESY 02-058\\
OUTP-02-23-P
\end{minipage}}\\
\vspace{2cm}
\bf Cosmic Microwave Background,\\ Matter-Antimatter Asymmetry\\
and Neutrino Masses}
\author{W.~Buchm\"uller, P. Di Bari\\
{\it Deutsches Elektronen-Synchrotron DESY, 22603 Hamburg, Germany}\\[5ex]
M.~Pl\"umacher\\
{\it Theoretical Physics, University of Oxford, 1 Keble Road,}\\
{\it Oxford, OX1 3NP, United Kingdom}
}
\maketitle

\thispagestyle{empty}

%\centerline{\date{\today}}

\begin{abstract}
\noindent
We study the implications of thermal leptogenesis for neutrino parameters.
Assuming that decays of $N_1$, the lightest of the heavy Majorana neutrinos,
initiate baryogenesis, we show that the final baryon asymmetry is
determined by only four parameters: the $C\!P$ asymmetry $\ve_1$, the heavy
neutrino mass $M_1$, the effective light neutrino mass $\mt$, and the quadratic
mean $\mb$ of the light neutrino masses. Imposing the CMB measurement of the
baryon asymmetry as constraint on the neutrino parameters,
we show, in a model independent way, that quasi-degenerate neutrinos are
incompatible with thermal leptogenesis. For maximal $C\!P$ asymmetry $\ve_1$, and
neutrino masses in the range from $(\D m^2_{sol})^{1/2}$ to $(\D m^2_{atm})^{1/2}$,
the baryogenesis temperature is $T_B = {\cal O}(10^{10})$~GeV.
\end{abstract}

\newpage

\section{Introduction}

The explanation of the cosmological baryon asymmetry is a challenge for particle
physics. In principle, already the standard model contains all necessary
ingredients, baryon number violation, $C$ and $C\!P$ violation, and also the required
departure from thermal equilibrium could be generated during the electroweak
phase transition \cite{rs96}. However, due to the lower bound on the Higgs boson
mass from LEP, electroweak baryogenesis is no longer a viable mechanism, except
for some supersymmetric extensions of the standard model \cite{qui01}.

A simple and elegant explanation of the observed baryon asymmetry is offered by
neutrino physics. During the past years data on atmospheric and solar neutrinos
have provided strong evidence for neutrino masses and mixings. In the seesaw
mechanism \cite{yan79} the smallness of these neutrino masses is naturally explained
by the mixing of the left-handed neutrinos with heavy Majorana neutrinos. Further,
the connection between baryon and lepton number in the high-temperature, symmetric
phase of the standard model due to rapid sphaleron transitions \cite{krs85} is
by now firmly established \cite{bmr00}. As in classical GUT baryogenesis \cite{kt90},
out-of-equilibrium decays of the heavy Majorana neutrinos can then generate a
lepton asymmetry which, by sphaleron processes, is partially transformed into a
baryon asymmetry \cite{fy86}.

A beautiful aspect of this `leptogenesis' mechanism is the connection between
the cosmological baryon asymmetry and neutrino properties. This connection is
established by standard kinetic calculations \cite{lut92,plu97}, very much like
in big bang nucleosynthesis \cite{kt90}, where light nuclei play the role analogous
to leptons in leptogenesis. The requirement of `successful baryogenesis', i.e.
the existence of neutrino masses and mixings for which the predicted and the
observed value of the baryon asymmetry are in agreement, constitutes a severe test
for models of neutrino masses, which has been extensively explored during the
past years \cite{bp00}.

On the experimental side, the precision of measurements of the baryon asymmetry
has significantly improved with the observation of the acoustic
peaks in the cosmic microwave background radiation (CMB). The BOOMERanG and DASI
experiments have measured the baryon asymmetry with a (1$\s$) standard error of
$\sim 15\%$ \cite{boo02,das02},
\begin{equation}
(\Omega_B\,h^2)^{CMB}=0.022^{+0.004}_{-0.003}\; ,
\end{equation}
Since the number of relic photons per comoving volume is very unlikely to have
changed after recombination, this is easily translated into a
measurement of the quantity $\eta_{B} = (n_B/n_{\gamma})$ at the present time,
\begin{equation}\label{obs}
\eta_{B0}^{CMB}=(6.0^{+1.1}_{-0.8})\times 10^{-10}\;.
\end{equation}
In the near future the MAP experiment will provide results of the first full
sub-degree sky survey of temperature anisotropies \cite{map02}.  The expected
(1$\s$) standard error on $\eta_{B0}$ is $\sim 10\%$,
using only CMB data, no polarization measurement and even allowing
for the presence of gravity wave perturbations \cite{jkx96}.
The PLANCK satellite, whose launch is planned for 2007 \cite{pla02},
should reduce this error to $\sim 1\%$.
If the polarization will be measured, and if it will be possible to add
extra CMB information on other cosmological parameters,
then the error may become even less than $1\%$ \cite{jkx96}.

In the following we shall calculate the baryon asymmetry by solving the Boltzmann
equations given in \cite{lut92,plu97}, assuming that the dominant contribution
is given by decays of $N_1$, the lightest of the heavy Majorana neutrinos.
This assumption is well justified in the case of a mass hierarchy among the
heavy neutrinos, i.e. $M_1 \ll M_2, M_3$, and it is also known to be a good
approximation, if $M_{2,3}-M_1 = {\cal O}(M_1)$ \cite{bdp02}. The case
$M_{2,3}-M_1 \ll M_1$ requires a special treatment. For some flavour structures
of the neutrino mass matrices it is also conceivable that the decays of the
heavier neutrinos $N_2$ or $N_3$ are the main source of the baryon asymmetry
\cite{nt01}. In our analysis we shall assume that quantum corrections \cite{bf00}
to the Boltzmann equations are small. So far, no detailed quantitative study
of this important question has been carried out.

As we shall see, within this framework one is left with only four parameters:
$M_1$, the CP asymmetry $\ve_1$, the effective neutrino mass $\mt$ and $\mb$,
the quadratic mean of the light neutrino masses. For each set of values of these
four parameters the Boltzmann equations yield a prediction for the baryon asymmetry.
A comparison with the observed value then defines an allowed region in the space
of neutrino parameters.

In addition to the neutrino parameters, there are also three quantities which
characterize the initial conditions: the initial temperature, the initial abundance
of heavy neutrinos and, of course, the initial baryon asymmetry. A detailed study
of the stability of the final baryon asymmetry under variations of these initial
conditions will be presented elsewhere \cite{bdp02}. In the following we shall
illustrate this dependence by presenting all results for two different choices of
the initial $N_1$ abundance, namely zero and thermal initial abundance. All values
in-between can be estimated by interpolation between these two cases. One can also
easily extrapolate the results to initial $N_1$ abundances higher than the thermal
one. Fortunately, for the most interesting range of $\mt$ the dependence on the
initial $N_1$ abundance turns out to be very small. Clearly, a theory of the very
early universe, like inflation, is needed to calculate the initial conditions for
baryogenesis.

The paper is organized as follows. In section~2 we recall the Boltzmann equations
which we then solve numerically. We also briefly discuss some approximations
underlying these equations. Section~3 deals with the theoretically allowed range
of the neutrino parameters, in particular the upper bound on the $C\!P$ asymmetry.
In section~4 we present our numerical results for the baryon asymmetry and discuss
the dependence on the neutrino parameters and on the choice of the initial condition.
We then investigate the constraints imposed by the CMB result on the neutrino
parameters. Our conclusions are given in section~5.

\section{Solutions of the Boltzmann equations}

The dynamical generation of a baryon asymmetry requires that the particle
interactions do not conserve baryon number, $C$ and $C\!P$. In leptogenesis
these conditions are realized by the couplings of the heavy Majorana
neutrinos $N_i$. Their decays can generate an asymmetry in the number of leptons
and antileptons, and therefore in $B-L$. The crucial departure from thermal
equilibrium is provided by the expansion of the universe. At temperatures
$T \sim {\cal O}(M_1)$ the abundance of heavy neutrinos exceeds the thermal
abundance due to their weak interactions with the thermal bath.

A quantitative description of this non-equilibrium process is obtained by means of
kinetic equations. The relevant processes in the thermal plasma are:
\begin{itemize}
\item $N_1$ decays ($D$) and inverse-decays ($I\!D$)
into leptons and Higgs bosons, $N_1\leftrightarrow \phi l$,
and into antileptons and anti-Higgs bosons, $N_1\leftrightarrow \bar{\phi} \bar l$;
\item $\D L=2$ scatterings mediated by the exchange of all heavy Majorana neutrinos,
$l \phi\leftrightarrow \bar{l} \bar{\phi}$ ($N$), and
$l l\leftrightarrow \bar{\phi}\bar{\phi}$,
$\bar{l} \bar{l}\leftrightarrow \phi \phi$ ($N,t$);
\item $\D L=1$ scatterings, $N_1 l(\bar{l}) \leftrightarrow
\bar{t}(t) q(\bar{q})$ ($\phi,s$)
and $N_1 t(\bar{t}) \leftrightarrow \bar{l}(l) q(\bar{q})$ ($\phi,t$);
\end{itemize}
in brackets we have indicated how the rates of these processes are labeled in the
following. In principle, one could also have additional processes, in particular
those which contribute to bring the heavy neutrinos initially into thermal
equilibrium \cite{plu97}. In the present, minimal framework we neglect such
interactions.

Since we are assuming that $N_1$ decays are the origin of lepton and baryon
asymmetries, the natural temperature scale is given by the mass $M_1$.
It is therefore convenient to measure temperature in units of $M_1$ and to
introduce the dimensionless variable $z = M_1/T$. For realistic values
$M_1\gg 100\,{\rm GeV}$, all standard model particles can be treated as massless,
and we shall assume that they are in thermal equilibrium.

The time evolution of a charge density or a number density $n_X$
depends on the microphysical processes in the thermal plasma as well as the
expansion of the universe. For the discussion of leptogenesis it is convenient to
consider instead of the number density $n_X$ the particle number $N_X$ in some
portion of comoving volume, which takes the effect of the
expansion automatically into account. We choose the comoving volume $R_{\star}(t)^3$
which contains one photon at time $t_{\star}$ before the onset of leptogenesis,
\begin{equation}
N_X(t) = n_X(t)\, R_{\star}(t)^3\;,
\end{equation}
with
\begin{equation}
R_{\star}(t_{\star}) = \left(n_\g^{eq}(t_{\star})\right)^{-1/3}\;,
\end{equation}
and therefore $N_{\g}(t_{\star})=1$. For a boson with $g_B$ degrees of freedom one
has $N_B(t_{\star}) = g_B/2$, whereas for a fermion with $g_F$ degrees of freedom
$N_F(t_{\star}) = 3g_F/8$. Alternatively, one may normalize the number density
to the entropy density $s$ and consider $Y_X = n_X/s$, as frequently done in
the literature. If entropy is conserved, both normalizations are related by a
constant. However, an inconvenient aspect of the quantity $Y_X$ is its dependence
on the entropy degrees of freedom, $g_S(t_{\star})/g_S(t_0)$, and on possible
entropy production between $t_{\star}$ and $t_0$.

The final baryon asymmetry is conveniently expressed in terms of the baryon-to-photon
ratio $\h_{B0} = n_B(t_0)/n_\g(t_0)$, to be compared with the CMB measurement
(\ref{obs}). The predicted value of $\h_{B0}$ is obtained from $N_B^0$ by
accounting for the dilution factor $f = N_\g(t_0) > 1$,
\begin{equation}
\h_{B0}\ =\ {1\over f}\ N_B^0 \;.
\end{equation}
In the simple case of constant entropy one has $f=g_S^{\star}/g_S^0$, with
$g_S^0 = 2 +21/11 \simeq 3.91$. Assuming at $t_{\star}$ the standard model degrees of
freedom with a single Majorana neutrino in addition, one obtains $g_S^{\star} = 434/4$,
and therefore $\h_{B0} \simeq 0.036 N_{B0}$.

In the decays of the heavy Majorana neutrinos an asymmetry in the number of lepton
doublets, and therefore in $B-L$, is generated. In the following we shall sum over
the three lepton numbers $L_e$, $L_\m$ and $L_\t$. Because of the large neutrino
mixings suggested by the solar and atmospheric neutrino anomalies, we expect this
to be a good approximation. A refined analysis can be performed along the lines
discussed in ref.~\cite{bcx00}. A related problem is the role of `spectator processes'
\cite{bp01} which change the naive sphaleron baryon-to-lepton conversion rate by a
factor ${\cal O}(1)$, since any generated asymmetry in lepton doublets is fast
distributed among many leptonic and baryonic degrees of freedom in the plasma.
In the following we shall ignore this uncertainty and
use the naive sphaleron conversion factor for $N_B/N_{B-L}$, which in the
standard model with one Higgs doublet is  $a = 28/79 \simeq 0.35$ \cite{ks88}.
The baryon-to-photon ratio today is then given by
\begin{equation}\label{etaB0}
\h_{B0}\ \simeq\ 0.013\ N_{B-L}^0\; .
\end{equation}

The Boltzmann equations for the time evolution of the number of heavy Majorana
neutrinos, $N_1$, and of $B-L$ number, $N_{B-L}$, are given by \cite{lut92,plu97},
\begin{eqnarray}
{dN_{N_1}\over dt} & = & -(\G_D+\G_S)\,(N_{N_1}-N_{N_1}^{\rm eq}) \label{lg10}\\
{dN_{B-L}\over dt} & = & -\ve_1\,\G_D\,(N_{N_1}-N_{N_1}^{\rm eq})
-\G_W\,N_{B-L} \label{lg20} \;.
\end{eqnarray}
Here the rate $\G_D$ accounts for decays and inverse decays ($z=M_1/T$),
\begin{equation}
\G_D= {1\over 8\p} \left(h^\dg h\right)_{11} M_1 {K_1(z)\over K_2(z)}\;,
\end{equation}
where $K_1$ and $K_2$ are Bessel functions, and $h$ is the Dirac neutrino Yukawa
matrix (cf.~section~3); the inverse decay rate is given by
$\G_{ID}=(n_{N_1}^{\rm eq}/n_{l})\,\G_D$.

The $N_1$ scattering rate involves processes with the Higgs field $\phi$ in
$t$- and $s$-channel,
\begin{equation}
\Gamma_S=2\Gamma_{\phi,t}^{(N_1)}+4\Gamma_{\phi,s}^{(N_1)}\;.
\end{equation}
Inverse decays, $\D L=1$ processes ($\G_{\phi,t}$, $\G_{\phi,s}$)
and $\D L=2$ processes ($\G_N$, $\G_{N,t}$) all contribute to the
washout rate,
\begin{equation}
\Gamma_W=\left({1\over 2}\G_{ID}+2\G_{\phi,t}^{(l)}+
\G_{\phi,s}^{(l)}{N_{N_1}\over N_{N_1}^{\rm eq}}\right)
+2\,\G_N^{(l)}+ 2\G_{N,t}^{(l)}
\end{equation}
The quantities $\G_{i}^{(X)}$ are thermally averaged reaction rates per particle $X$.
They are related by $\G_i^{(X)}=\g_i/n_{X}^{\rm eq}$ to the reaction densities
$\g_i$ \cite{lut92} which are obtained from the reduced cross sections
$\hat{\s}_i(s/M_1^2)$,
\begin{equation}
\g_{(i)}(z)={M_1^4\over 64\pi^4}{1\over z}
\int_{(m_a^2+m_b^2)/M_1^2}^{\infty}\,dx
\hat{\s}_{(i)}(x)\,\sqrt{x}\,K_1(z\,\sqrt{x})\;,
\end{equation}
where $m_a$ and $m_b$ are the masses of the two particles in the initial state.
Our calculations are based on the reduced cross sections given in ref.~\cite{plu98}.

An important part of our analysis is an improved treatment of the $\D L=2$ processes
which involve the heavy Majorana neutrinos $N_i$, $i=1\ldots 3$, as intermediate states.
The reduced cross sections $\hat{\s}_N$ and $\hat{\s}_{(N,t)}$ read,
\begin{equation}\label{gammaN}
\hat{\s}_{N(N,t)}= {1\over 2\p}
\left[\sum_{i} (h^\dg h)_{ii}^2\,f_{ii}^{N(N,t)} (x)
+ \sum_{i<j}{\cal R}e (h^\dg h)_{ij}^2\,f_{ij}^{N(N,t)} (x)\right],
\end{equation}
with
\begin{eqnarray}
f_{ii}^N(x) &=& 1+{a_j\over D_j(x)}+{x\,a_j\over 2D_j^2(x)}-{a_j\over x}
\left[1+{x+a_j\over D_j}\right]\,\ln \left(1+{x\over a_j}\right) \; ,\\
f_{ij}^{N}(x) &=& \sqrt{a_i\,a_j}\left[
{1\over D_i(x)}+{1\over D_j(x)}+{x\over D_i(x)\,D_j(x)}\right. \NO \\
&&\hspace{1.5cm}+\left(1+{a_i\over x}\right)\left({2\over a_j-a_i}-{1\over D_j(x)}\right)\,
\ln\left(1+{x\over a_i}\right) \NO \\
&& \left.\hspace{1.5cm}\left(1+{a_j\over x}\right)\left({2\over a_i-a_j}-{1\over D_i(x)}\right)\,
\ln\left(1+{x\over a_j}\right)\right]\; ,  \\
f_{ii}^{N,t}(x) &=& {x\over x+a_j}+{a_j\over x+2a_j}
\ln\left(1+{x\over a_j} \right)\; , \\
f_{ij}^{N,t}(x) &=& {\sqrt{a_i\,a_j} \over (a_i-a_j)(x+a_i+a_j)}
\left[(2x+3a_i+a_j)\ln\left(1+{x\over a_j}\right)\right. \NO\\
&& \hspace{3cm}\left. -(2x+3a_j+a_i)\ln\left(1+{x\over a_i}\right)\right]\; .
\end{eqnarray}
Here $a_j\equiv M^2_j/M_1^2$, and $1/D_i(x)\equiv (x-a_i)/[(x-a_i)^2+a_i c_i]$
is the off-shell part of the $N_i$ propagator with
$c_i =  a_i(h^{\dg}h)_{ii}^2/(8\pi)^2$.

The sum $\g_{N}+\g_{N,t}$ is conveniently separated in two parts.
The first part comes from the resonance contribution $\propto x/D_1^2$,
which is highly peaked around $x=1$. This term is easily evaluated analytically
in the zero-width limit,
\begin{equation}\label{on}
\g_{N}^{\rm res} = {M_1^4\over 64 \p^3} (h^\dg h)_{11} {1\over z} K_1(z)\; .
\end{equation}
For typical values of $(h^\dg h)_{11}$ and $M_1$ the resonance contribution dominates
in the temperature range from $T\simeq M_1$ ($z\simeq 1$) down to $T\simeq 0.1 M_1$
($z \simeq 10$). The remaining part is dominant at low temperatures, $z > 10$.
For $z \gg 1$ the main contribution to the integrals $\g_{N}$ and $\g_{N,t}$ comes
from the region $x\ll 1$. Here the scattering amplitudes with $N_i$ exchange
are proportional to the light neutrino mass matrix $m_\n$. To leading
order in $x$ for $\hat{\s}_N$ and $\hat{\s}_{(N,t)}$, one finds
\begin{equation}\label{off}
\g_N(z\gg 1) \simeq \g_{(N,t)}(z\gg 1)
\simeq {3 M_1^6\over 8\p^5 v^4} {1\over z^6} {\rm tr}\left(m_\n^\dg m_\n\right)\;.
\end{equation}
For hierarchical neutrinos
$\mb^2 = {\rm tr}\left(m_\n^\dg m_\n\right) \simeq \D m^2_{atm}$,
whereas for quasi-degenerate neutrinos one has $\mb^2 \simeq 3 m^2_i$, with
$m_1 \simeq m_2 \simeq m_3 \gg \D m^2_{atm}$.

Since all rates are expressed as functions of $z$, it is convenient to replace time
by $z=M_1/T$ in the eqs.~(\ref{lg10}) and (\ref{lg20}). This change of variables
introduces the Hubble parameter since $dt/dz = 1/(Hz)$, with
$H \simeq 1.66\,\sqrt{g_{\r}}(M_1^2/ M_{\rm Pl})/z^{2}$ where $g_{\r}$ is the
number of energy degrees of freedom at $t_{\star}$. Neglecting the small variation
of the number of degrees of freedom during leptogenesis one obtains for the standard
model with one right-handed neutrino $g_\r = 434/4$.

The kinetic equations for leptogenesis now read,
\begin{eqnarray}\label{ke}
{dN_{N_1}\over dz} & = & -(D+S)\,(N_{N_1}-N_{N_1}^{\rm eq}) \;, \label{lg1} \\
{dN_{B-L}\over dz} & = & -\ve_1\,D\,(N_{N_1}-N_{N_1}^{\rm eq})-W\,N_{B-L} \;,\label{lg2}
\end{eqnarray}
where we have defined $(D,S,W) = (\Gamma_{D},\Gamma_S,\Gamma_W)/(H\,z)$.
In order to understand the dependence of the solutions on the neutrino parameters,
it is crucial to note that the rates $\G_D$, $\G_S$ and also $\G_W$, except for
the contribution  $\D \G_W = \G_N^{(l)} - \G^{(l)}_{N,res}+\G_{N,t}^{(l)}$,
are all proportional to $(h^\dg h)_{11}$. The rescaled rates in eqs.~(\ref{lg1}) and
(\ref{lg2}) are therefore dimensionless functions of $z$, proportional to
\begin{equation}\label{scaling}
D,\;  S,\;  W-\D W \; \propto \,  {M_{\rm Pl}\mt \over v^2} \;, \quad
\D W \, \propto \,  {M_{\rm Pl}M_1\, \mb^2\over v^4}  \; ;
\end{equation}
the effective neutrino mass $\mt$ \cite{plu97} is given by
\begin{equation}
\mt = {(m_D^\dg m_D)_{11} \over M_1} \; ,
\end{equation}
where $m_D$ is the Dirac neutrino mass matrix (cf.~section~3).
Eq.~(\ref{scaling}) implies that, as long as $\D W$ can be neglected, the generated
lepton asymmetry is independent of $M_1$.

Since $\D W$ increases with $M_1$, it does become important at large values of $M_1$.
At $z\gg 1$, one easily obtains from (\ref{off}),
\begin{equation}\label{dwl}
\Delta W (z>>1)= \alpha\,{M_1\,\mb^2\over z^2} \;,
\end{equation}
with $\a^{-1} = \zeta(3)\pi^3\,g_l\,v^4\,H(z=1)/M_1^2$.

For $z < 1$, $\D W$ is no longer proportional to $M_1\mb^2$, but depends
on the heavy neutrino masses $M_i$ and on the specific structure of the $h$ matrix.
An approximate upper bound $\G_W^+$ on $\G_W$ is given by the sum of the resonance
contribution $\G^{(l)}_{N,res}$ and $\D \G(z\gg 1)$ taken at all values of $z$. In this
way the relativistic suppression of $N_1$ exchange at high temperatures is underestimated.
Analogously, a lower limit $\G_W^-$ can be obtained by treating in the off-shell part
$\D \G$ the heavy neutrinos $N_2$ and $N_3$ kinematically like $N_1$, which overestimates
the relativistic suppression of $N_{2,3}$ exchange at high temperatures. The two branches
$\G_W^{\pm}$ are shown in figs.~1a-4a. As expected, the uncertainty of $\G_W(z)$
at small $z$ is not important for the final baryon asymmetry in most cases. Numerically,
we find that is negligible for heavy neutrino masses
$M_1 \lesssim 10^{13}\ {\rm GeV}\ (0.1\ {\rm eV}/\mb)^2$.

Successful leptogenesis requires a departure from thermal equilibrium for the
decaying heavy Majorana neutrinos. Furthermore, at the same time, washout processes
must not be in thermal equilibrium. The corresponding naive out-of-equilibrium
conditions are $\left.(\G_D + \G_S)\right|_{z=1} < \left.H\right|_{z=1}$ and
$\left.\G_W\right|_{z=1} < \left.H\right|_{z=1}$. These conditions are fulfilled
for a typical set of parameters $M_1 = 10^{10}$~GeV, $\mt = 10^{-3}$~eV,
$\mb = 0.05$~eV, for which the three rates and the Hubble parameter are shown
in fig.~1a. The generation of the $B-L$ asymmetry for these parameters is shown
in fig.~1b for $|\ve_1|=10^{-6}$. The figure also demonstrates that the Yukawa
interactions are strong enough to bring the heavy neutrinos into thermal equilibrium.
The resulting asymmetry is in accord with observation.

Increasing $\mt$ by two orders of magnitude to
$\mt = 10^{-1}$~eV increases all rates while leaving the Hubble
parameter unchanged. As fig.~2a shows, the out-of-equilibrium condition for
$N_1$ decays is now no longer fulfilled, and the final $B-L$ asymmetry is reduced
by two orders of magnitude. Increasing $M_1$ to $10^{15}$~GeV increases $\D W$
by five orders of magnitude, which now dominates the
shape of the washout rate $\G_W$ (fig.~3a). Although the Yukawa interactions are
strong enough to bring the heavy neutrinos into thermal equilibrium, the
washout rate is now so large that the final $B-L$ asymmetry is reduced by three
orders of magnitude. Finally, reducing $\mt$ to
$10^{-5}$~eV while keeping $M_1 = 10^{10}$~GeV fixed
(fig.~4a), one becomes dependent on the initial conditions. The Yukawa interactions
are no longer strong enough to bring the heavy neutrinos into thermal
equilibrium (fig.~4b). Starting from zero initial abundance, $N_{N_1}^{in} = 0$,
the final $B-L$ asymmetry is reduced by one order of magnitude compared to
fig.~1b. On the other hand, assuming initially a thermal distribution, which may
have been generated by other interactions, the final asymmetry is enhanced by
one order of magnitude compared to fig.~1b. This is the `way-out-of-equilibrium'
case, where washout effects can be neglected, and the final $B-L$ asymmetry is
given by $3\ve_1/4$.

\section{The CP asymmetry}

Consider now the standard model with right-handed neutrinos. The neutrino
masses are obtained from the lagrangian,
\begin{equation}
{\cal L}_m
= h_{ij}\Bar{l}_{Li}\n_{Rj}\phi
 + {1\over 2} M_{ij}\Bar{\n}^c_{Ri}\n_{Rj} + h.c.\;.
\end{equation}
Here $M$ is the Majorana mass matrix of the right-handed neutrinos, and the
Yukawa couplings $h$ yield the Dirac neutrino mass matrix $m_D = h v$
after spontaneous symmetry breaking, $v = \VEV {\phi}$. We work in the
mass eigenstate basis of the right-handed neutrinos where $M$ is diagonal
with real and positive eigenvalues $M_1 \leq M_2 \leq M_3$. The seesaw
mechanism \cite{yan79} then yields for the light neutrino mass matrix,
\begin{equation}
m_{\nu}= - m_D{1\over M}m_D^T \;, \label{seesaw}
\end{equation}
where higher order terms in $1/M$ have been neglected.

The mass matrix $m_\n$ can again be diagonalized by a unitary matrix $U^{(\n)}$,
\begin{equation}\label{mdiag}
U^{(\n)\dg} m_{\n} U^{(\n)*}  = - \left(\begin{array}{ccc}
    m_1  & 0  & 0\\
    0   &  m_2   & 0  \\
    0  & 0    & m_3
    \end{array}\right) \equiv - D_m \;,
\end{equation}
with real and positive eigenvalues satisfying $m_1 \leq m_2 \leq m_3$
Inserting eq.~(\ref{seesaw}) one finds,
\begin{equation}
v^2 U^{(\n)\dg}h D_M^{-1} h^T U^{(\n)*} = D_m \;,
\end{equation}
which means that
\begin{equation}\label{ortho}
\O = v D_m^{-1/2} U^{(\n)\dg} h D_M^{-1/2}
\end{equation}
is an orthogonal matrix, $\O \O^T = I$ \cite{ci01}. This also implies
Im($\O^T \O$)$_{11} = 0$, from which one immediately obtains
\begin{equation}\label{rel}
{1\over m_1} \mbox{Im}\left(U^{(\n)\dg}h\right)^2_{11} =
- \sum_{i\neq 1} {1\over m_i} \mbox{Im}\left(U^{(\n)\dg}h\right)^2_{i1}\;.
\end{equation}

The $C\!P$ asymmetries in the decays of the heavy Majorana neutrinos arise
at one-loop order from the interference of the tree level amplitude
with vertex and self-energy corrections \cite{fps95,crv96,bp98}. In the
following, we shall restrict ourselves to the case of hierarchical heavy
neutrinos, i.e. $M_1 \ll M_2, M_3$. The $C\!P$ asymmetry $\ve_1$ for the
decay of $N_1$ is then easily obtained by first integrating out the heavier
fields $N_2$ and $N_3$. From the tree amplitude and the effective
lepton-Higgs interaction one then obtains the useful expression \cite{bf00},
\begin{equation}
\ve_1 \simeq - {3\over 16\pi} {M_1\over (h^\dg h)_{11}}
 \mbox{Im}\left(h^\dg h {1\over M} h^T h^*\right)_{11}\;,
\end{equation}
where corrections ${\cal O}(M_1/M_{2,3})$ have been neglected.
Using eqs.~(\ref{mdiag}) and (\ref{rel}) one easily derives,
\begin{equation}\label{nice}
\ve_1 = {3\over 16\p} {M_1 \over v^2} \sum_{i \neq 1} {\D m_{i1}^2\over m_i}
{\mbox{Im}\left(\tilde{h}^2_{i1}\right) \over
\left(\tilde{h}^\dg \tilde{h}\right)_{11}}\; ,
\end{equation}
where $\D m_{i1}^2 = m_i^2 - m_1^2$, and
\begin{equation}\label{yeigen}
\tilde{h} = U^{(\n)\dg} h
\end{equation}
is the matrix of Yukawa couplings in the mass eigenstate basis of light and
heavy Majorana neutrinos. In the interesting case $M_2 - M_1 = {\cal O}(M_1)$
the $C\!P$ asymmetry $\ve_1$ is enhanced by a factor ${\cal O}(1)$. However,
one then has to study the decays of both heavy neutrinos, $N_1$ and $N_2$
\cite{bdp02}.

In the case of hierarchical neutrinos,
$m_3\simeq (\Delta m^2_{atm})^{1/2}\gg m_2\simeq (\Delta m^2_{sol})^{1/2}\gg m_1$,
eq.~(\ref{nice}) yields an upper bound on the $C\!P$ asymmetry \cite{di02}
\begin{equation}\label{bgen}
|\ve_1| \leq \,{3 \over 16 \p} {M_1 (\D m^2_{atm})^{1/2} \over v^2} \;,
\end{equation}
where we have neglected the term  $\propto \D m^2_{sol}$.
For an inverted hierarchy $m_3\sim m_2\simeq(\Delta m^2_{atm})^{1/2}\gg m_1$
and $m_3^2-m_2^2=\Delta m^2_{sol}$ both terms contributing to eq.~(\ref{nice})
are approximately equal to $(\D m^2_{atm})^{1/2}$, and the upper bound is
larger by a factor two.
%In the case of normal hierarchical neutrinos
%the two mass squared differences can be identified with $\D m^2_{sol}$
%and $\D m^2_{atm}$ observed in solar and atmospheric neutrino oscillations.
%From eq.~(\ref{nice}), neglecting
%the term $\propto \D m^2_{sol}$, one then obtains the upper bound \cite{di02}:
%\begin{equation}\label{bgen}
%|\ve_1| \leq \,{3 \over 16 \p} {M_1 (\D m^2_{atm})^{1/2} \over v^2} \;.
%\end{equation}
%For inverted hierarchical neutrinos  the two terms are both approximately
%equal to $(\D m^2_{atm})^{1/2}$ and the upper bound is twice larger.
Further, for Yukawa couplings with $|h_{ij}| \leq |h_{33}| = {\cal O}(1)$, one has
$M_3 \sim v^2/m_3$. The maximal $C\!P$ asymmetry is then a measure of the
hierarchy among the heavy Majorana neutrinos,
\begin{equation}
|\ve_1|^{\rm max} \sim 0.1\, {M_1 \over M_3}\; .
\end{equation}
It is remarkable that the upper bound (\ref{bgen}) is frequently saturated in models
with hierarchical neutrino masses (see, e.g., \cite{bcx00}, \cite{bp96}-\cite{bs02}).
One reason
is that in the leptonic mixing matrix $U = U^{(e)\dg}U^{(\n)}$ all elements, except
$U_{e3}$, are known to be ${\cal O}(1)$. This is often
explained by the structure of $U^{(\n)}$, whereas $U^{(e)}$ has small
off-diagonal elements. The Yukawa matrix $\tilde{h}_\n = U^{(\n)\dg} h_\n$ has
then naturally large off-diagonal elements, even if $h_\n$ is almost diagonal.
For hierarchical neutrinos the upper bound on the $C\!P$ asymmetry is then easily
saturated.

In terms of the quadratic mean $\mb/\sqrt{3}$, with
\begin{equation}
\mb = \sqrt{m_1^2 + m_2^2 + m_3^2}\; ,
\end{equation}
the bound (\ref{bgen}) may be written as
\begin{equation}\label{hier}
|\ve_1| \leq \,{3\over 16 \p} {M_1 \mb \over v^2} \;.
\end{equation}
For quasi-degenerate neutrinos, where
$m_1 \simeq m_2 \simeq m_3 \simeq \mb/\sqrt{3} \gg (\D m^2_{atm})^{1/2}$,
one gets the stronger bound
\begin{equation}\label{deg}
|\ve_1| \leq {3\sqrt{3}\over 16\p} {M_1 \D m^2_{atm} \over v^2 \mb} \;.
\end{equation}
There is no well defined boundary between hierarchical and quasi-degenerate
neutrinos. In our analysis we shall choose $\mb = 1$~eV, such that
$\mb/\sqrt{3} \simeq 0.58\,{\rm eV} \gg \sqrt{\D m^2_{atm}} \simeq 0.05\,{\rm eV}$.
The interesting case where all neutrinos have masses
$m_i = {\cal O}(0.1\,{\rm eV})$ will be discussed elsewhere \cite{bdp02}.

What is the allowed range of the masses $\mt$ and $M_1$ which are crucial for
leptogenesis? Assuming Yukawa couplings $|\tilde{h}_{ij}| \leq 1$, one has
\begin{equation}
M_1 \leq M_3 \sim {v^2 \over m_3}\; .
\end{equation}
The effective neutrino mass $\mt$ is given by (cf.~(\ref{ortho})),
\begin{equation}
\mt = {v^2\over M_1}\sum_i |\tilde{h}_{i1}^2|
= \sum_i m_i |\O_{i1}^2| \; .
\end{equation}
From the orthogonality of $\O$ one obtains the lower bound \cite{fhy02},
\begin{equation}\label{mt1lbound}
\mt \geq m_1 \sum_i |\O_{i1}^2| \geq m_1 \sum_i {\rm Re}(\O_{i1}^2) = m_1\; .
\end{equation}
The orthogonality condition for $\O$ reads explicitely,
\begin{equation}
{\rm Re}(\O^T\O)_{11} =
{v^2\over M_1}\sum_i {1\over m_i}{\rm Re}(\tilde{h}_{i1}^2) = 1\; .
\end{equation}
Unless there are strong cancellations due to phase relations between different
matrix elements, one then obtains
\begin{equation}
\mt \leq m_3 {v^2\over M_1}\sum_i {1\over m_i}|\tilde{h}_{i1}^2|
 \sim m_3 {v^2\over M_1}\sum_i {1\over m_i}{\rm Re}(\tilde{h}_{i1}^2) = m_3\; .
\end{equation}
In our analysis we shall therefore emphasize the range $m_1 \leq \mt \leq m_3$.

\section{Constraints on neutrino parameters}

In this section we shall compare the predicted baryon asymmetry with the value
measured from CMB (cf.~(\ref{obs})), using the relation (\ref{etaB0}).
Incorporating also the bound on the CP asymmetry described in the
previous section will enable us to derive an allowed region for the three parameters
$M_1$, $\mt$ and, very interestingly, $\mb$. Before describing the numerical
results it is useful to discuss some general properties of the solutions of the
kinetic equations.

The abundance $N_{N_1}(z)$ of the heavy Majorana neutrinos is independent of the
asymmetry $N_{B-L}$. Thus, for a given solution $N_{N_1}(z)$, or equivalently
$\Delta (z) = N_{N_1}(z)-N_{N_1}^{\rm eq}(z)$, the solution for the asymmetry
$N_{B-L}(z)$ reads,
\begin{equation}
N_{B-L}(z)=N_{B-L}(z_{\rm in})\,e^{-\int_{z_{\rm in}}^{z}\,dz'\,W(z')}
+{3\over 4}\,\ve_1\,\k(z;\mt,M_1,\mb)\;,
\end{equation}
where $z_{\rm in} \ll 1$, and the efficiency factor $\k$ is given by
\begin{equation}\label{kappa}
\k(z)= {4\over 3} \int_{z_{\rm in}}^{z}\,dz'\,D(z')\Delta(z')
\,e^{-\int_{z'}^{z}dz''\,W(z'')}\;.
\end{equation}
Since we assume that the initial asymmetry is zero, $N_{B-L}(z)$ is proportional to the
CP asymmetry $\ve_1$.

A interesting, limiting case is $W=0$ and $N_{N_1}^{\rm eq}=0$, i.e. $\D=N_{N_1}$, which
corresponds to far out-of-equilibrium decays with vanishing washout. The final baryon
asymmetry then takes the form
\begin{equation}
N_{B-L}^0 = \ve_1 \, N_{N_1}(z_d)\;,
\end{equation}
where $z_d$ is some temperature at which the heavy neutrinos are out of equilibrium
and have not yet decayed. For an initial thermal abundance one then
obtains $N_{B-L}^0 = 3\ve_1/4$,
and therefore $\k_0 = 1$. Within thermal leptogenesis this is the maximal asymmetry,
and one always has $\k_0 \leq 1$.

In special parameter regimes it is possible to obtain explicit analytical
expressions for the efficiency factor $\k(z;\mt,M_1,\mb)$ \cite{bcx00}.
For small $\mt$ and $\mb$ washout effects are
small. Starting from $N_1^{\rm in}=0$, the final asymmetry is then proportional to
the generated $N_1$ abundance. From the reaction densities given in \cite{plu98}, one
easily obtains at $z \ll 1$: $D \propto M_{pl}\mt z^2/v^2$ and
$S \propto M_{pl}\mt /v^2$. Hence, one obtains from eq.~(\ref{lg1}),
\begin{equation}\label{low}
\k \propto N_{N_1}(z_d) \propto {M_{pl}\mt \over v^2}\; .
\end{equation}
Also interesting is the case of large $\mt$ and small $M_1$, where the washout is
dominated by the $W-\Delta W$ term. At large temperatures, $z<1$, the thermal
$N_1$ abundance is then quickly reached. At small temperatures, $z>1$, the
$N_1$ abundance is reduced by the $\D L=1$ scatterings and by the decays.
The reaction densities $\g_{S,D}$ decrease exponentially,
and the efficiency factor is determined by
the $N_1$ abundance at freeze-out $\bar{z}$ \cite{bcx00},
\begin{equation}\label{high}
\k \propto  N_{N_1}(\bar{z}) \propto {1\over \g_{S,D}} \propto {1\over \mt}\; .
\end{equation}
For intermediate temperatures, i.e. $z \sim 1$, the Boltzmann equations have to
be solved numerically.

\subsection{Numerical results}

In fig.~5 we have plotted\footnote{All numerical results have been crosschecked by
two independent codes. Figs.~1-4 and 5b-7b are based on one code; the
other code has been used for figs.~5a-7a.} the predicted present baryon asymmetry
as a function of the parameter $\mt$ for different values of $M_1$, assuming a
typical value of the $C\!P$ asymmetry, $\ve_1=-10^{-6}$. We have calculated
the baryon asymmetry for two values of $\mb$. In fig.~5a,
$\mb=0.05\,{\rm eV}\simeq \sqrt{\D m^2_{\rm atm}}$,
which corresponds to the case of hierarchical neutrinos; $\mb =1\,{\rm eV}$
in fig.~5b, which represents the case of quasi-degenerate neutrinos with
all light neutrino masses approximately equal,
$m_i = 1/\sqrt{3}\,{\rm eV} \simeq 0.58\,{\rm eV}$. We have performed the
calculations both for an initially vanishing $N_1$ abundance (thick lines) and
for a thermal initial $N_1$ abundance, $N_{N_1}^{\rm in}=3/4$ (thin lines).

Following \cite{bcx00}, we show in fig.~6 the dependence of the baryon asymmetry
on the parameters $\mt$ and $M_1$ by means of iso-$\k_0$ curves of the efficiency
parameter. Figs.~6a and 6b correspond to the cases $\mb = 0.05\,{\rm eV}$ and
$\mb = 1\,{\rm eV}$, respectively. For vanishing initial $N_1$ abundance the
enclosed domains have a finite extension in $\mt$; for thermal initial abundance
there is no boundary at small $\mt$.

Figures~5 and 6 clearly show the existence of two different regimes: the
domain of `small' $N_1$ masses,
$M_1 < 10^{13}\,{\rm GeV} (0.1\,{\rm eV}/\mb)^2$,
and the domain of `large' $N_1$ masses,
$M_1 > 10^{13}\,{\rm GeV} (0.1\,{\rm eV}/\mb)^2$. The dependence of the boundary
on $\mb$ is determined by the behaviour of the non-resonant washout rate,
$\D W \propto M_1 \mb^2$. Note, that in previous studies of the washout effects
\cite{bp00} the obtained $M_1$ dependence was a result of an assumed behaviour of
$\D W$ at $z \gg 1$. As our discussion in section~2 shows, such an assumption is
unnecessary, since the behaviour of $\D W$ is governed by  $M_1 \mb^2$.

\subsubsection{Small $M_1$ regime}

In this case the non-resonant part $\D W$ of the washout rate is negligible
in first approximation, and there is only a small dependence on $M_1$ and $\mb$
in the final asymmetry. Hence, the efficiency factor $\kappa_0$ depends
approximately only on the parameter $\mt$. From fig.~5 one reads off that in
the case of zero initial $N_1$ abundance the final asymmetry reaches its
maximum at $\mt^{\rm peak}\simeq 6\times 10^{-4}\,{\rm eV}$.
We can then express the baryon asymmetry in the following form,
\begin{equation}
\eta_{B0}= 1.5\times 10^{-9}\,{|\ve_1|\over 10^{-6}}\,
{\k_0(\mt)\over \k_0^{\rm peak}}\;,
\end{equation}
where $\k_0^{\rm peak} \equiv \k_0(\mt^{\rm peak})\simeq 0.16$ is the maximal
efficiency factor $\k_0$ obtainable for zero initial $N_1$ abundance.

The behaviour of the efficiency factor for small and large values of $\mt$ is
known analytically (cf.~(\ref{low}), (\ref{high})). Approximately,
one has $\k_0 \propto\mt$ for $\mt \ll \mt^{\rm peak}$ , and
$\k_{0}\propto 1/\mt$ for $\mt \gg \mt^{\rm peak}$. A good fit for
$\k_0(\mt)$ is given by the following expression,
\begin{equation}\label{fit}
\k_0(\mt)\simeq 0.24 \left(x_-\,e^{-x_-}+x_+e^{-x_+} \right)\;,
\end{equation}
with
\begin{equation}
x_{\pm}=\left(\mt \over \widetilde{m}_{\pm} \right)^{\mp 1 - \a}\;,
\end{equation}
and $\widetilde{m}_{-} = 3.5 \times 10^{-4}\,{\rm eV}$,
$\widetilde{m}_{+} = 8.3 \times 10^{-4}\,{\rm eV}$, $\a = 0.1$.
The corresponding baryon asymmetry is shown in fig.~5a; the fit is optimal
for $M_1=10^{8}\,{\rm GeV}$.

In case of thermal initial $N_1$ abundance the maximal efficiency factor is
obtained in the unphysical limit $\mt \rightarrow 0$, where the heavy neutrinos
decouple completely from the thermal bath. One then has
\begin{equation}
\eta_{B0}^{\rm max}\simeq 0.96\times 10^{-2}\,\ve_1 \;.
\end{equation}
On the other hand, for sufficiently strong coupling of the heavy neutrinos
to the thermal bath, i.e. $\mt \gtrsim \mt^{\rm peak}$, there is no
dependence on the initial neutrino abundance, and the two cases
$N_{N_1}=0$ and $N_{N_1}^{\rm in}=3/4$ give the same final baryon asymmetry.
In this case possible asymmetries generated in the decays of the heavier
neutrinos $N_2$ and $N_3$ have also no effect on the final baryon asymmetry.
Note, that in comparison with the analytical results of ref.~\cite{bcx00}, we obtain an
efficiency factor $\k$ which is about three times smaller.

\subsubsection{Large $M_1$ regime}

In this regime the washout rate $\D W\propto M_1\,\mb^2$ dominates and the
iso-$\k_0$ curves in the $(\mt,M_1)$-plane tend to become independent of
$\mt$ with increasing $M_1$. However, some dependence on $\mt$ remains due to the
effect of $D$ and $S$ (cf.~(\ref{lg1}),(\ref{lg2})) on the $N_1$ abundance.
This is why the curves never become
exactly horizontal. Note, that the curves tend to converge in such a way that
a small variation of $M_1$ leads to a large variation of the final asymmetry. In
this region any small change in the kinetic equations can have large effects on
the final asymmetry. On the other hand, a curve corresponding to a given value of
the final asymmetry, e.g. $\eta_{B0}^{CMB}$, is rather insensitive to small changes
of the kinetic equations. Thus in this region the uncertainty of $\D W$ at small
values of $z$ is, fortunately, not important. We therefore conclude that,
within our framework of thermal leptogenesis, we have obtained a description of
the baryon asymmetry in terms of just four neutrino parameters: $\ve_1$, $M_1$,
$\mt$ and $\mb$.

\subsection{CMB constraint}

The CMB constraint on neutrino parameters is given by the requirement
\begin{equation}\label{bound}
\eta_{B0}\simeq 0.96\times 10^{-2}\,
|\ve_1|\,\kappa_0(\mt,M_1,\mb)=\eta_{B0}^{CMB}\;.
\end{equation}
In the case $N_{N_1}^{\rm in}=0$, the substitution of the efficiency factor
by its maximum $\k_0^{\rm peak}$ yields an important lower limit on the
$C\!P$ asymmetry $\ve_1$,
\begin{equation}\label{CMBb}
|\ve_1|\ \gtrsim\ 4.0 \times 10^{-7}\
\left(\eta_{B0}^{CMB}\over 6\times10^{-10}\right)\
\gtrsim\  2.4\ \times 10^{-7}\;.
\end{equation}
Here the last inequality corresponds to the (3$\s$) lower limit for $\h_{B0}^{CMB}$
(cf.~(\ref{obs})).
For small $M_1$, the CMB bound becomes a determination of $\mt$ as function of the
CP asymmetry. There are obviously two solutions: $\mt^- < \mt^{\rm peak}$
and $\mt^+ > \mt^{\rm peak}$. From the fit (\ref{fit}) one easily finds
$\mt^{\mp}$ for $C\!P$ asymmetries sufficiently far above the bound (\ref{CMBb}),
\begin{eqnarray}
\mt^- &\simeq& 7.9\times 10^{-5}\,{\rm eV}\,
\left({\eta_{B0}^{CMB}\over 6\times 10^{-10}}
{{10^{-6}\over |\ve_1|}}\right)^{1.1}\; , \\
\mt^+ &\simeq& 2.8 \times 10^{-3}\,{\rm eV}\,
\left({6\times 10^{-10}\over \eta_{B0}^{CMB}}
{{|\ve_1|\over 10^{-6}}}\right)^{0.9}\; .
\end{eqnarray}
As $\ve_1$ decreases to the lower bound (\ref{CMBb}), $\mt^-$ and $\mt^+$
approach $\mt^{\rm peak} \simeq 6\times 10^{-4}$~eV.

In the case of a thermal initial $N_1$ abundance the bound (\ref{CMBb}) gets
relaxed by a factor $1/k_0^{\rm peak}\simeq 6.4$, and one obtains the usual
bound for $\k_0 = 1$ in eq.~(\ref{bound}),
\begin{equation}\label{CMBb2}
|\ve_1|\ \gtrsim\ 6.3 \times 10^{-8}\
\left(\eta_{B0}^{CMB}\over 6\times 10^{-10}\right)\ \gtrsim\ 3.8\times 10^{-8}\; .
\end{equation}
This bound can only be reached for $\mt \ll \mt^{\rm peak}$, i.e. very small
values which are not easily obtained in models of neutrino masses. Note, that
$\mt > m_1$, the smallest neutrino mass eigenvalue \cite{fhy02} and that
$\mt^{\rm peak} \sim 0.1 \sqrt{\D m^2_{sol}}$. Further, one has to worry about
the production mechanism of a large initial abundance of extremely weakly
coupled heavy neutrinos.

\subsection{CMB constraint plus CP bound}

Until now we have treated the CP asymmetry $\ve_1$ as an independent parameter.
However, as discussed in section~3, for hierarchical as well as
quasi-degenerate neutrinos, the CP asymmetry satisfies an upper bound
$|\ve_1|<\ve(M_1,\mb)$ (cf.~eqs.~(\ref{hier}), (\ref{deg})). Together
with the CMB constraint (\ref{bound}) this yields the following restriction
on the space of parameters $\mt$, $M_1$, and $\mb$,
\begin{equation}\label{constraint}
\eta_{B0}^{\rm max}(\mt,M_1,\mb)\simeq
0.96\times 10^{-2}\,\ve(M_1,\mb)\,\k_0(\mt,M_1,\mb) \gtrsim  \eta^{CMB}_{B0}\; .
\end{equation}

For hierarchical neutrinos eq.~(\ref{hier}) yields the upper bound on the
$C\!P$ asymmetry,
\begin{equation}\label{max1}
|\ve_1| < 1\,\times 10^{-6}\,\left({M_1\over 10^{10}\,{\rm GeV}}\right)
\left({\D m^2_{atm}\over 2.5 \times 10^{-3}{\rm eV^2}}\right)^{1/2}\;
\end{equation}
Combining this with the CMB constraint (\ref{CMBb}) yields a lower bound on
$M_1$ \cite{di02}. For zero (thermal) initial $N_1$ abundance one obtains,
\begin{equation}\label{lowM}
M_1 \gtrsim 2.4 \, (0.4) \times 10^{9}\,{\rm GeV}
\left({2.5\times 10^{-3}{\rm eV^2}\over \D m^2_{atm}}\right)^{1/2}\;.
\end{equation}
In the case of inverted hierarchy the upper bound on the $C\!P$ asymmetry
is twice as large and therefore the lower bound on $M_1$ is twice as small.

In the case of quasi-degenerate neutrinos eq.~(\ref{deg}) implies a stronger
bound on the $C\!P$ asymmetry,
\begin{equation}\label{max2}
|\ve_1| < 0.9\times 10^{-7}\,\left({M_1\over 10^{10}\,{\rm GeV}}\right)
\left({1 {\rm eV}\over \mb}\right)
\left({\D m^2_{atm}\over 2.5 \times 10^{-3}{\rm eV^2}}\right)\; .
\end{equation}
The corresponding lower bound on $M_1$ reads for zero (thermal) initial
$N_1$ abundance,
\begin{equation}
M_1\gtrsim 2.7 \, (0.4) \times 10^{10}\,{\rm GeV}
\left({\mb \over 1 {\rm eV}}\right)
\left({2.5 \times 10^{-3}{\rm eV^2} \over \D m^2_{atm}}\right)\; .
\end{equation}

The bounds (\ref{max1}) and (\ref{max2}) on the $C\!P$ asymmetry seem to suggest
that by increasing $M_1$ any value $|\ve_1|<1$ can be reached. One may therefore
expect it to be rather easy to satisfy the CMB constraint (\ref{bound}).
However, in the regime of large $M_1$ the efficiency factor $\k_0(\mt,M_1,\mb)$
is exponentially suppressed. This dominates the linear increase of
$|\ve_1|^{\rm max}$ with $M_1$ and leads to an upper limit on $M_1$.

This situation is illustrated in figs.~7a and 7b, where iso-$\eta_{B0}^{\rm max}$
curves in the $(\mt, M_1)$-plane are shown for $\mb = 0.05$~eV and $\mb = 1$~eV,
respectively. In the first case we have assumed a normal hierarchy.
In the second case we have used $|\ve_1|<10^{-7}M_1/10^{10}$GeV
corresponding to $\D m^2_{atm} \simeq 2.9\times 10^{-3}{\rm eV^2}$.
For values $\mt$ and $M_1$ enclosed by these curves the baryon
asymmetry $\eta_{B0}^{CMB}$ can be obtained. In the case of $N_{N_1}^{\rm in}=0$,
the allowed regions are closed domains; for $N_{N_1}^{\rm in}=3/4$ there is no
lower limit on $\mt$.

As discussed in section~3, the effective neutrino mass $\mt$ is bounded from
below by $m_1$, the mass of the lightest neutrino, and also likely to be
smaller than $m_3$, the mass of the heaviest neutrino.
For quasi-degenerate neutrinos this implies $\mt \simeq \mb/\sqrt{3}$. This
value is indicated by the vertical line in fig.~7b from which one reaches the
conclusion that quasi-degenerate neutrinos are incompatible with leptogenesis.
A similar conclusion was reached in ref.~\cite{fhy02} based on results of \cite{bp00},
assuming zero initial $N_1$ abundance. As our analysis shows quasi-degenerate
neutrinos are strongly disfavoured by thermal leptogenesis
for all possible $C\!P$ asymmetries and initial conditions. Possible ways
to evade this conclusion
are a resonant enhancement of the $C\!P$ asymmetry in the case of heavy neutrino
mass differences of order the decay widths, $|M_{2,3} - M_1|={\cal O}(\G_i)$
\cite{pil99}, or a completely non-thermal leptogenesis \cite{fhy02}.

In the case of  hierarchical neutrinos there is a lower bound on $M_1$ as function
of $\mt$, which can be read off from fig.~7a in the case of a normal
hierarchy, while for inverted hierarchy this bound is twice as small.
No upper bound stronger than
$M_3 \sim 10^{15}$~GeV exists. For thermal initial conditions all values of
$\mt < m_3$ are allowed.

Comparing figs.~7a and 7b, it is evident that a more stringent upper limit
than 1~eV exists for $\mb$. The determination of this precise bound goes beyond
the goal of this paper and is left for future work \cite{bdp02}.

Any point in the space $(\eta_{B0},\tilde{m}_1,M_1)$ which lies below the
surface $\eta_{B0}^{\rm max}$ and above the plane $\eta_{B0}^{CMB}$ represents
a possible set of neutrino parameters with some value $|\ve_1|\leq |\ve_1|^{\rm max}$.
As discussed in section~3, in many models one has $|\ve_1| \sim |\ve_1|^{\rm max}$.
The CMB bound then yields a precise relation between $\mt$ and $M_1$. In particular,
for the interesting range
$0.1\,{\rm eV}\gtrsim \mt\gtrsim \mt^{\rm peak} \simeq 6\times 10^{-4}\,{\rm eV}$,
one has the simple relation,
\begin{equation} \label{M1}
M_{1}\simeq 3\times 10^{10}\,{\rm GeV}\,\left({\mt\over 0.01 {\rm eV}}\right)\;.
\end{equation}
This corresponds to the scenario originally proposed in \cite{bp96}, with
$\mt \sim \sqrt{\D m^2_{sol}} \sim 0.005 {\rm eV}$, where we have assumed the
LMA solution \cite{bgp01}. The corresponding baryogenesis temperature is
$T_B \sim M_1 \sim 10^{10}$~GeV.

\section{Conclusions}

We have studied in some detail the minimal version of thermal leptogenesis. In this
framework the decays of $N_1$, the lightest of the heavy Majorana neutrinos, are the
source of the baryon asymmetry, and only Yukawa interactions generate the initial
heavy neutrino abundance; further, the connection between light and heavy neutrinos
is given by the seesaw mechanism. The final baryon asymmetry is then determined
by only four parameters: the $C\!P$ asymmetry $\ve_1$, the heavy neutrino mass
$M_1$, the effective light neutrino mass $\mt$ and the quadratic mean $\mb$ of the
light neutrino masses.

The constraint from the cosmic microwave background on the baryon asymmetry
strongly restricts the allowed range of neutrino parameters. For small values of
$M_1$, the efficiency factor $\k$ depends only on $\mt$. Together with the upper
bound on the $C\!P$ asymmetry, this yields an important lower bound on $M_1$ and
therefore on the baryogenesis temperature, $T_B \sim M_1 \geq 2.4\times 10^9$~GeV.

For large values of $M_1$, washout effects, proportional to $M_1\mb^2$, become
important. This leads to the conclusion that the case of quasi-degenerate neutrinos,
$\mb \gg \sqrt{\D m^2_{atm}}$, is incompatible with thermal leptogenesis. On the
other hand, in the case of hierarchical neutrinos, the full range of light and
heavy neutrino masses considered in grand unified theories is viable. The intermediate
regime, $m_i = {\cal O}(0.1 {\rm eV})$, remains to be studied in detail.

Global fits of solar and atmospheric neutrino data indicate large mixing angles for
neutrino oscillations in both cases. With respect to neutrino masses, leptogenesis
favours the possible case of a mild hierarchy, $m_2 \simeq \sqrt{\D m^2_{sol}}
\simeq 0.1\ \sqrt{\D m^2_{atm}} \simeq 0.1\ m_3$. Due to the large mixing angles,
the a priori possibility $m_1 \ll m_2$ appears unlikely. With
$m_1 \leq \mt \lesssim m_3$, one then arrives at the conclusion
$\mt \sim 0.1 \sqrt{\D m^2_{atm}}$. Furthermore, in many of the recently
discussed models for neutrino masses the $C\!P$ asymmetry $\ve_1$ is close to its
upper bound. This finally determines the heavy neutrino mass $M_1$ and the
baryogenesis temperature to be $T_B \sim M_1 = {\cal O}(10^{10}\ {\rm GeV})$.

The unfortunate prediction of thermal leptogenesis is that the search for absolute
neutrino masses in tritium $\b$-decay and CMB combined with Large Scale Structure
will be unsuccessful in the near future. Our present bound,
$\sum_i m_i < \sqrt{3}\ {\rm eV}$, can still be significantly improved \cite{bdp02}.
However, the search for neutrinoless double $\b$-decay with a sensitivity
$|m_{ee}|={\cal O}(\sqrt{\D m^2_{atm}})$ \cite{gen01} may very well be successful. \\

\noindent
\textbf{Acknowledgments}

P.D.B. was supported by the Alexander von Humboldt foundation;
M.P.~was supported by the EU network ``Supersymmetry and the
Early Universe'' under contract no.~HPRN-CT-2000-00152. We wish to thank
S.~Davidson, S.~Huber, A.~Ibarra, H.~B.~Nielsen, Q.~Shafi and Y. Takanishi
for useful discussions.

\newpage
\psfig{file=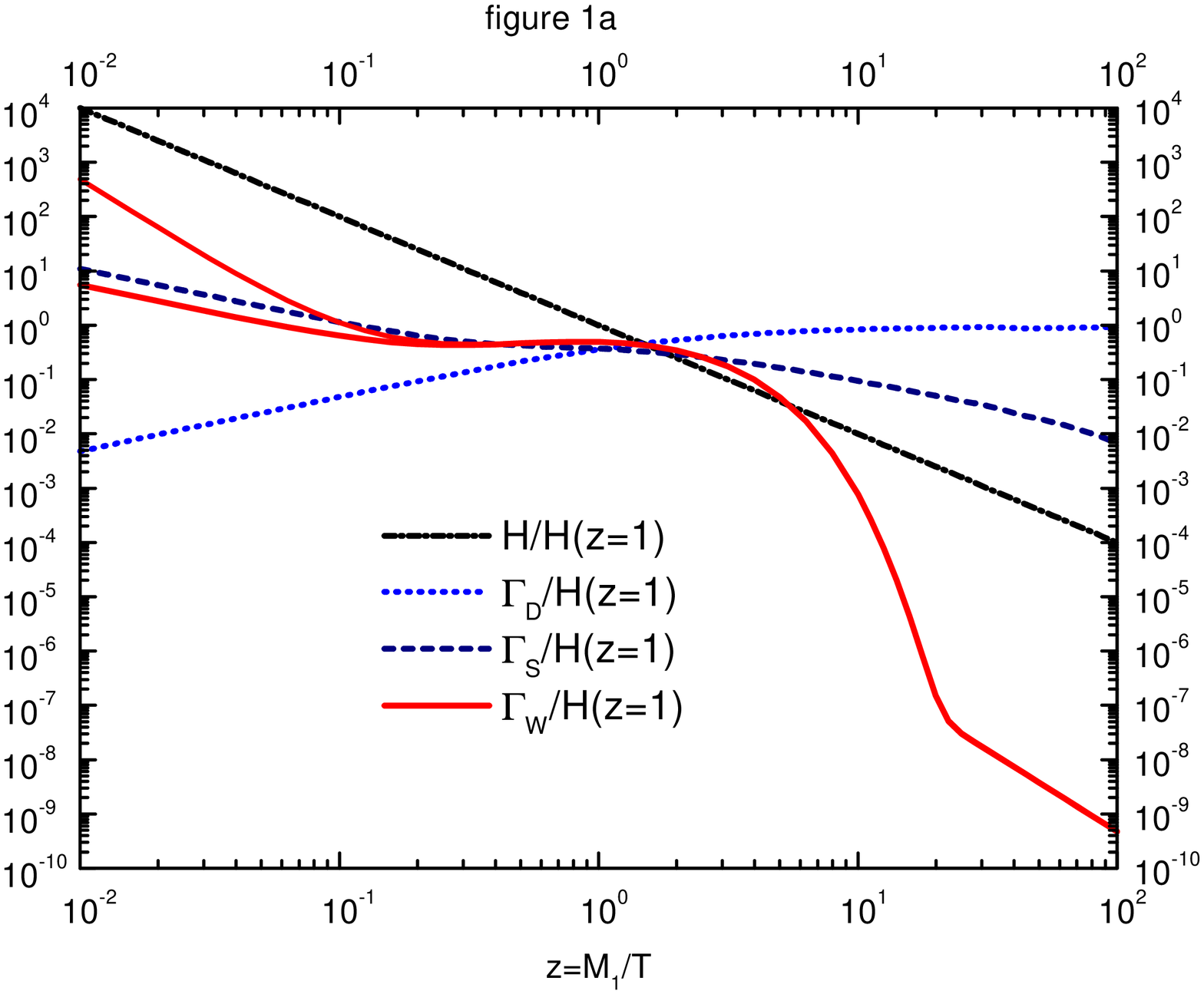,width=14cm}

\psfig{file=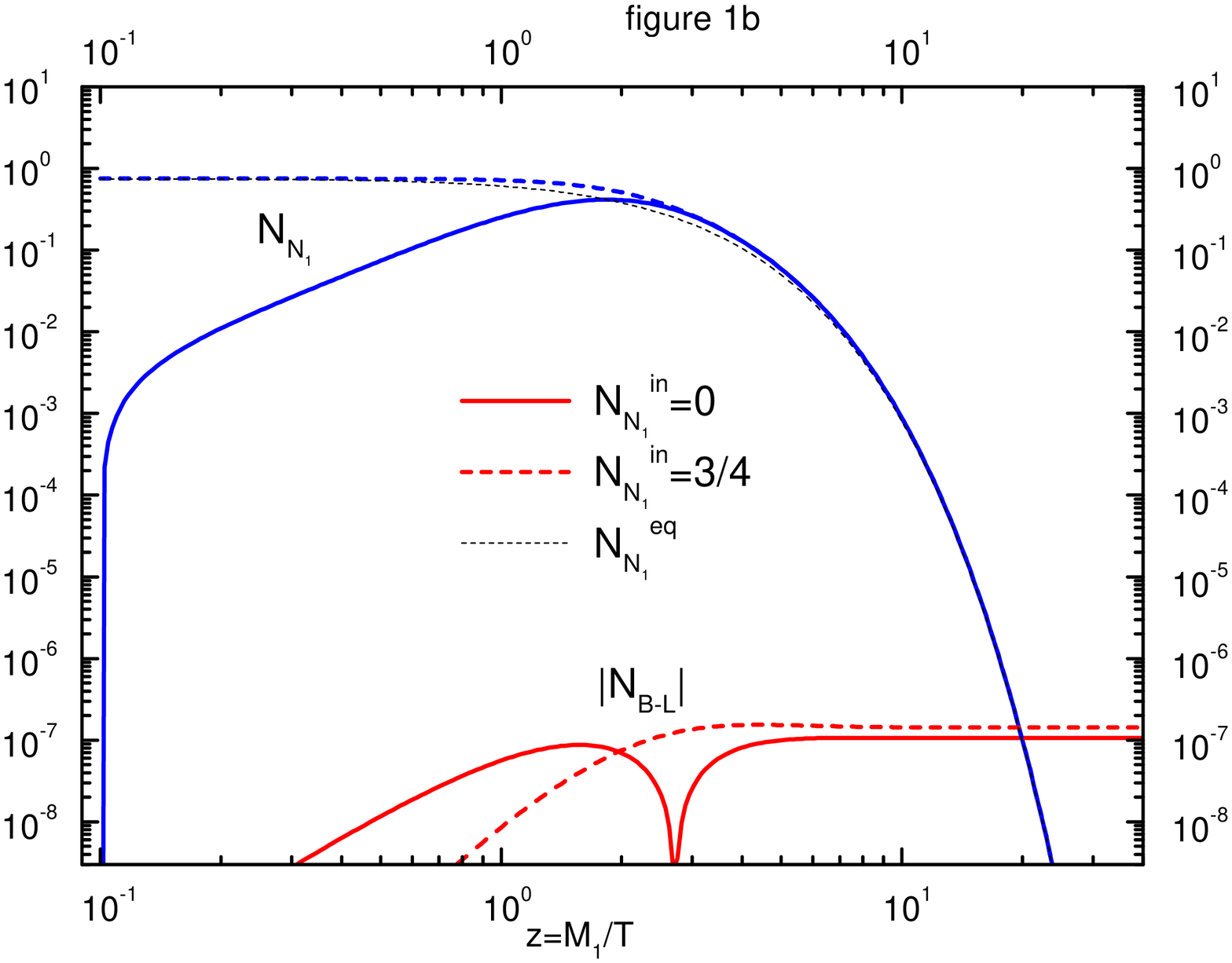,width=14cm}

{\bf Figure 1.} $M_1=10^{10}\,{\rm GeV}$, $\mt=10^{-3}\,{\rm eV}$.
(a) Rates normalized to the
expansion rate at $z=1$. The two branches for $\G_W$ at small
$z$ correspond to the upper (lower) bounds $\G_W^+$ ($\G_W^-$) (see text).
(b) Evolution of the $N_1$ abundance and the $B-L$ asymmetry
for $\ve_1=-10^{-6}$ and $\mb=0.05\,{\rm eV}$,
both zero and thermal initial $N_1$ abundance.

\newpage
\psfig{file=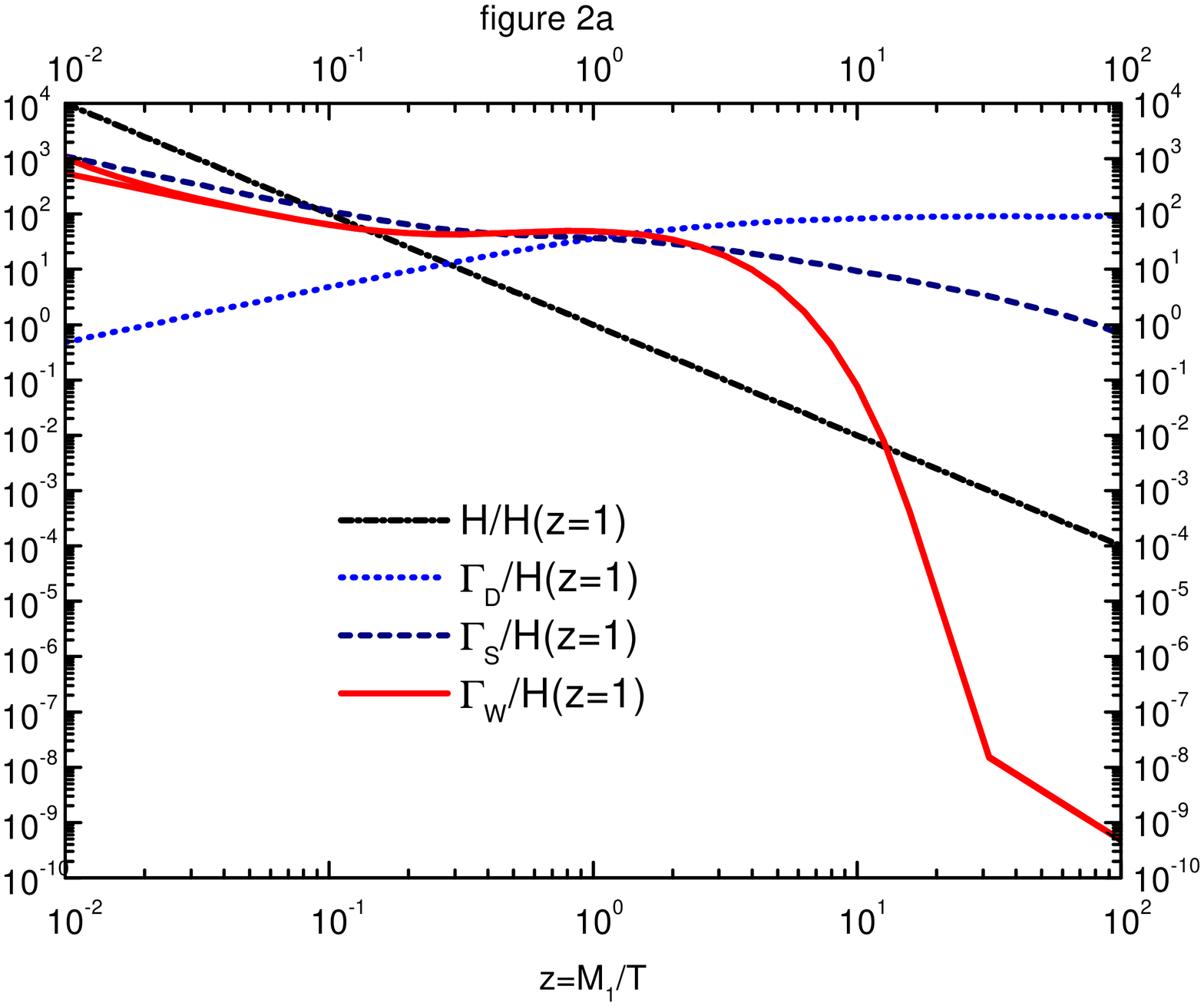,width=14cm}

\psfig{file=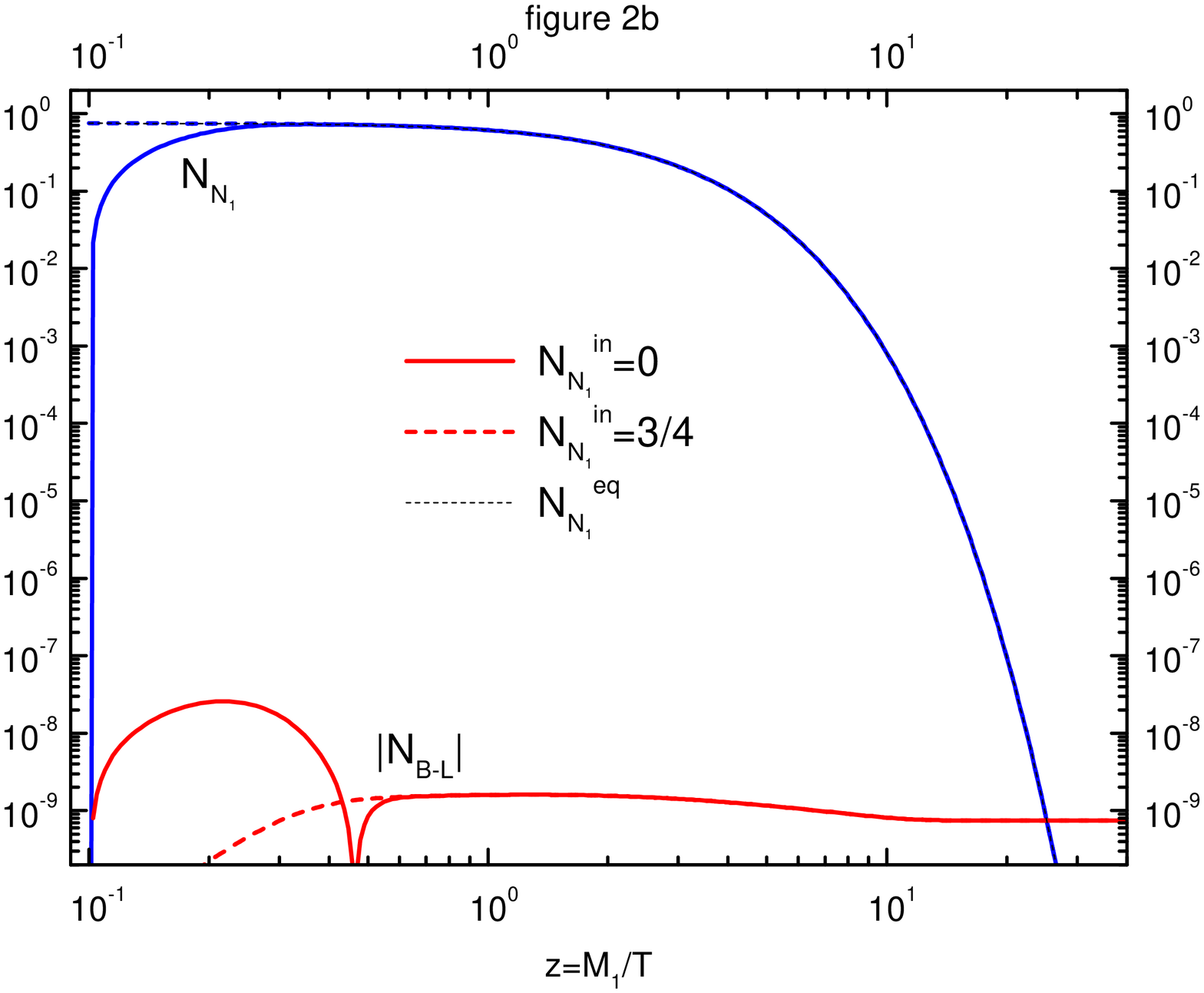,width=14cm}

{\bf Figure 2.} (a) Rates and (b) evolution of $N_1$ abundance $B-L$ asymmetry;
parameters as in figure~1, except $\mt=10^{-1}\ {\rm eV}$.

%$M_1=10^{10}\,{\rm GeV}$,
%$\tilde{m}_1=10^{-1}\,{\rm eV}$. (a) Rates.
%(b) Evolution of the $N_1$ abundance and of the $B-L$ asymmetry
%for $\epsilon_1=-10^{-6}$ and $\bar{m}=0.05\,{\rm eV}$.

\newpage
\psfig{file=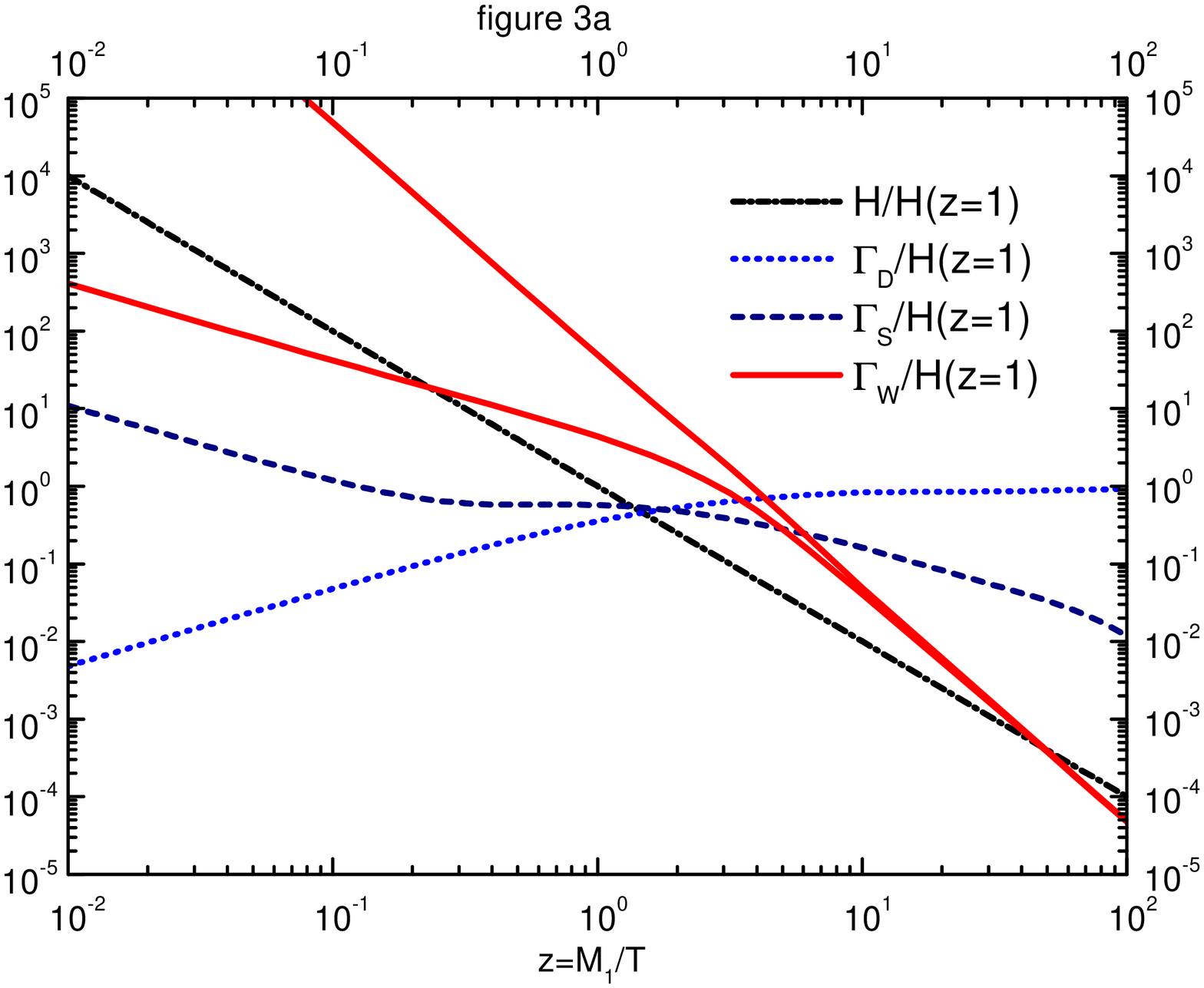,width=14cm}

\psfig{file=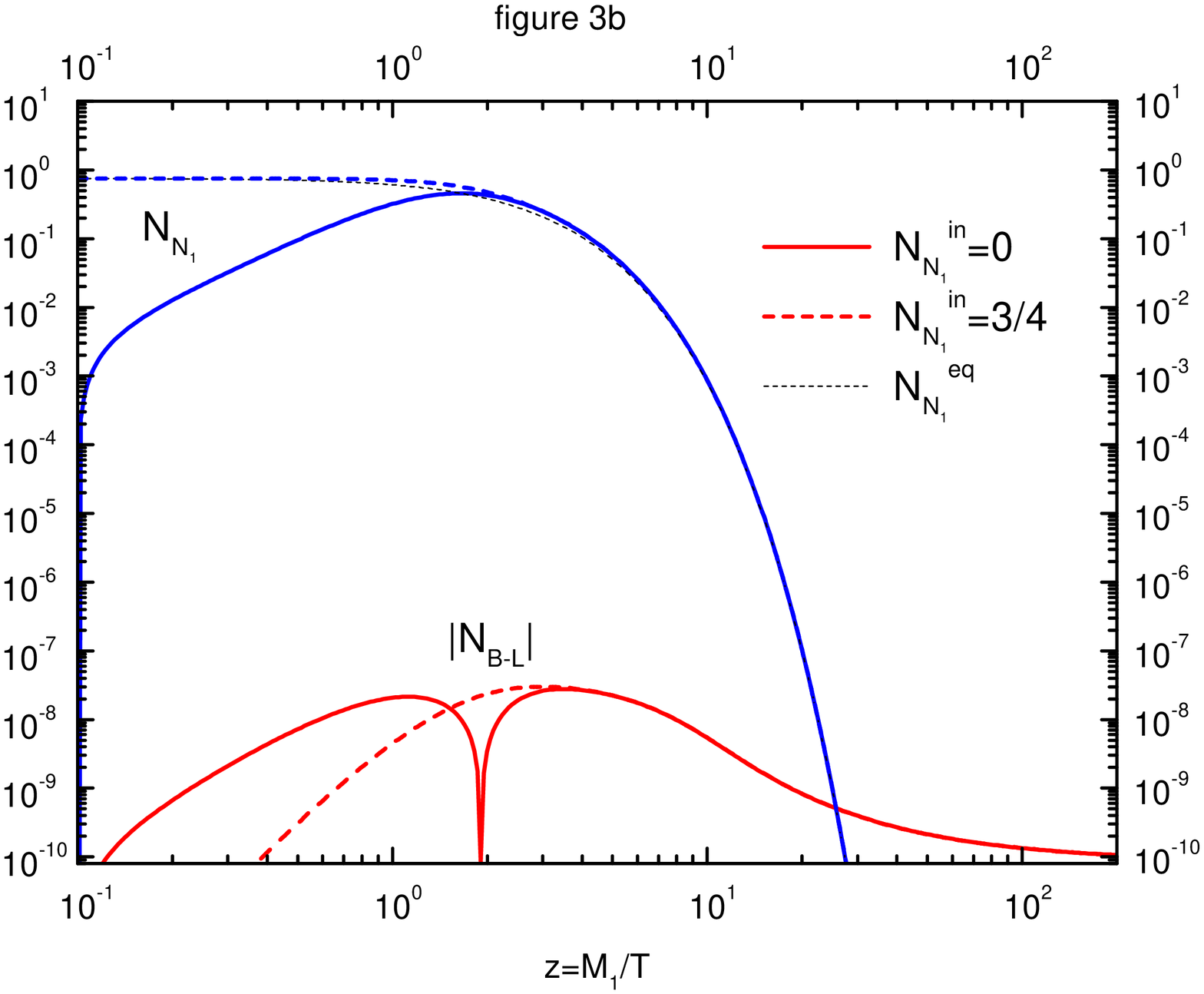,width=14cm}

{\bf Figure 3.} (a) Rates and (b) evolution of $N_1$ abundance $B-L$ asymmetry;
parameters as in figure~1, except $M_1=10^{15}\ {\rm GeV}$.
%$\tilde{m}_1=10^{-3}\,{\rm eV}$. (a) Rates.
%(b) Evolution of the $N_1$ abundance and of the $B-L$ asymmetry
%for $\epsilon_1=-10^{-6}$ and $\bar{m}=0.05\,{\rm eV}$.

\newpage
\psfig{file=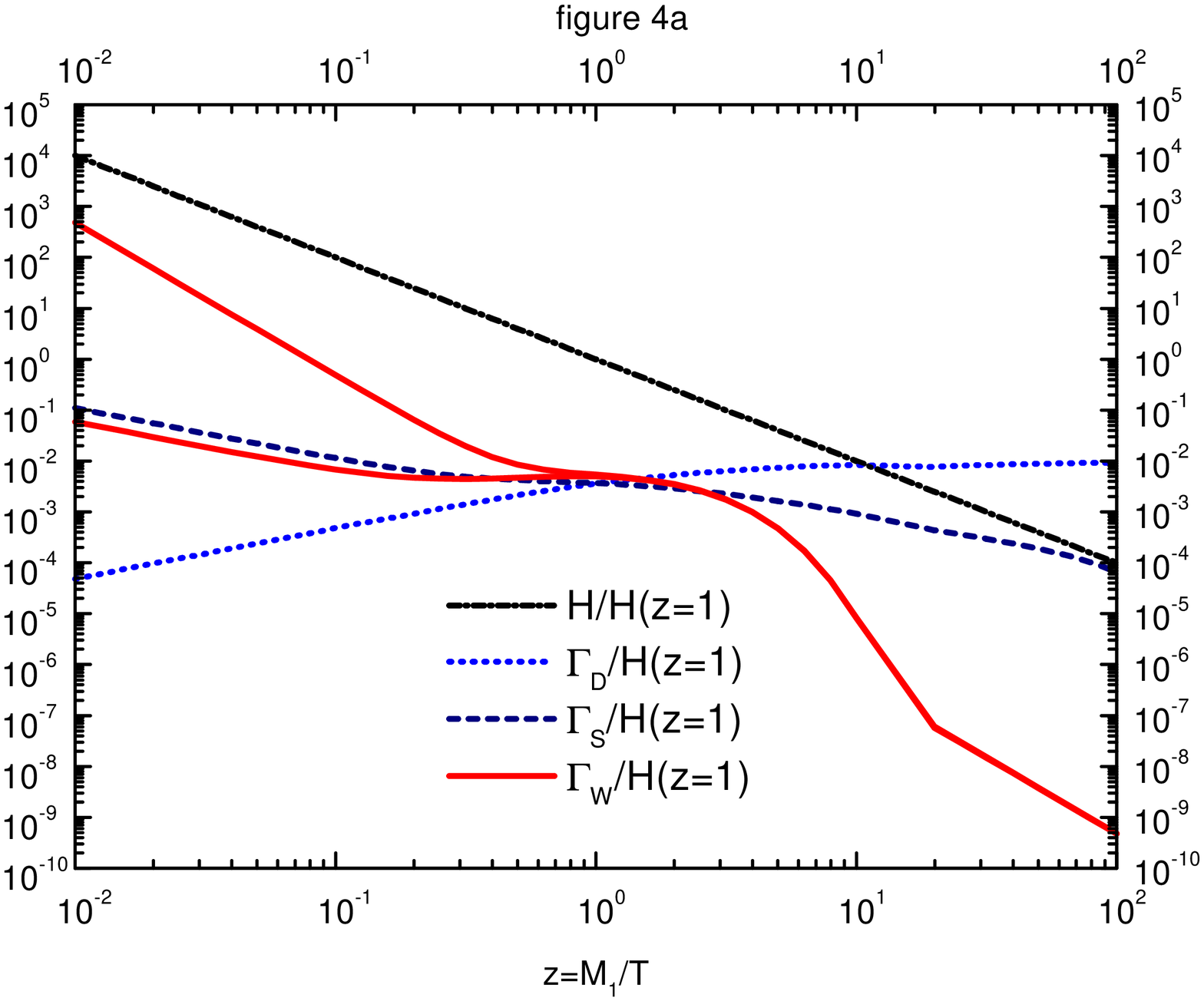,width=14cm}

\psfig{file=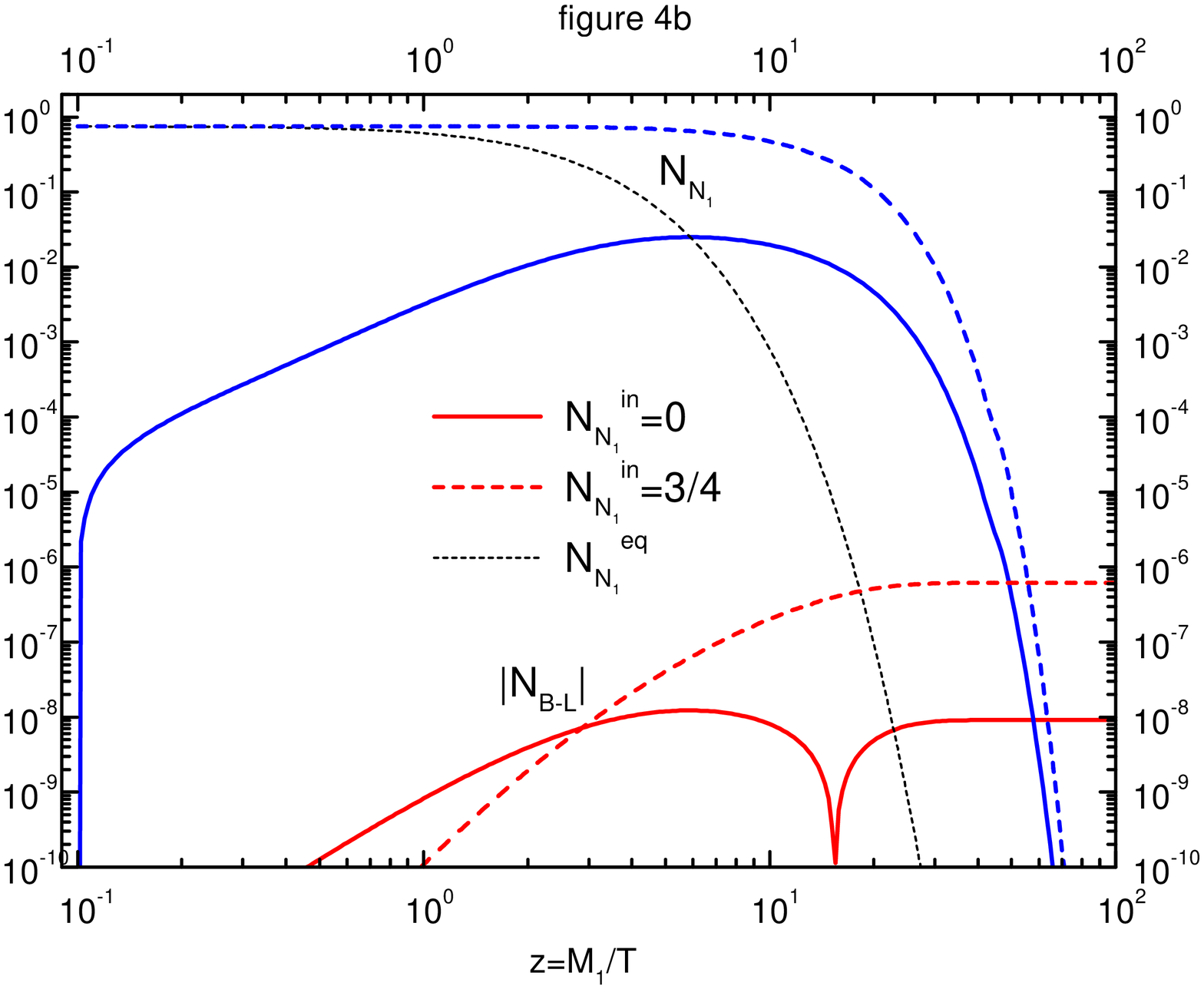,width=14cm}

{\bf Figure 4.} (a) Rates and (b) evolution of $N_1$ abundance $B-L$ asymmetry;
parameters as in figure~1, except $\mt=10^{-5}\ {\rm eV}$.
%$M_1=10^{10}\,{\rm GeV}$,
%$\tilde{m}_1=10^{-5}\,{\rm eV}$. (a) Rates.
%(b) Evolution of the $N_1$ abundance and of the $B-L$ asymmetry
%for $\epsilon_1=-10^{-6}$ and $\bar{m}=0.05\,{\rm eV}$.

\newpage
\hspace{11mm}
\psfig{file=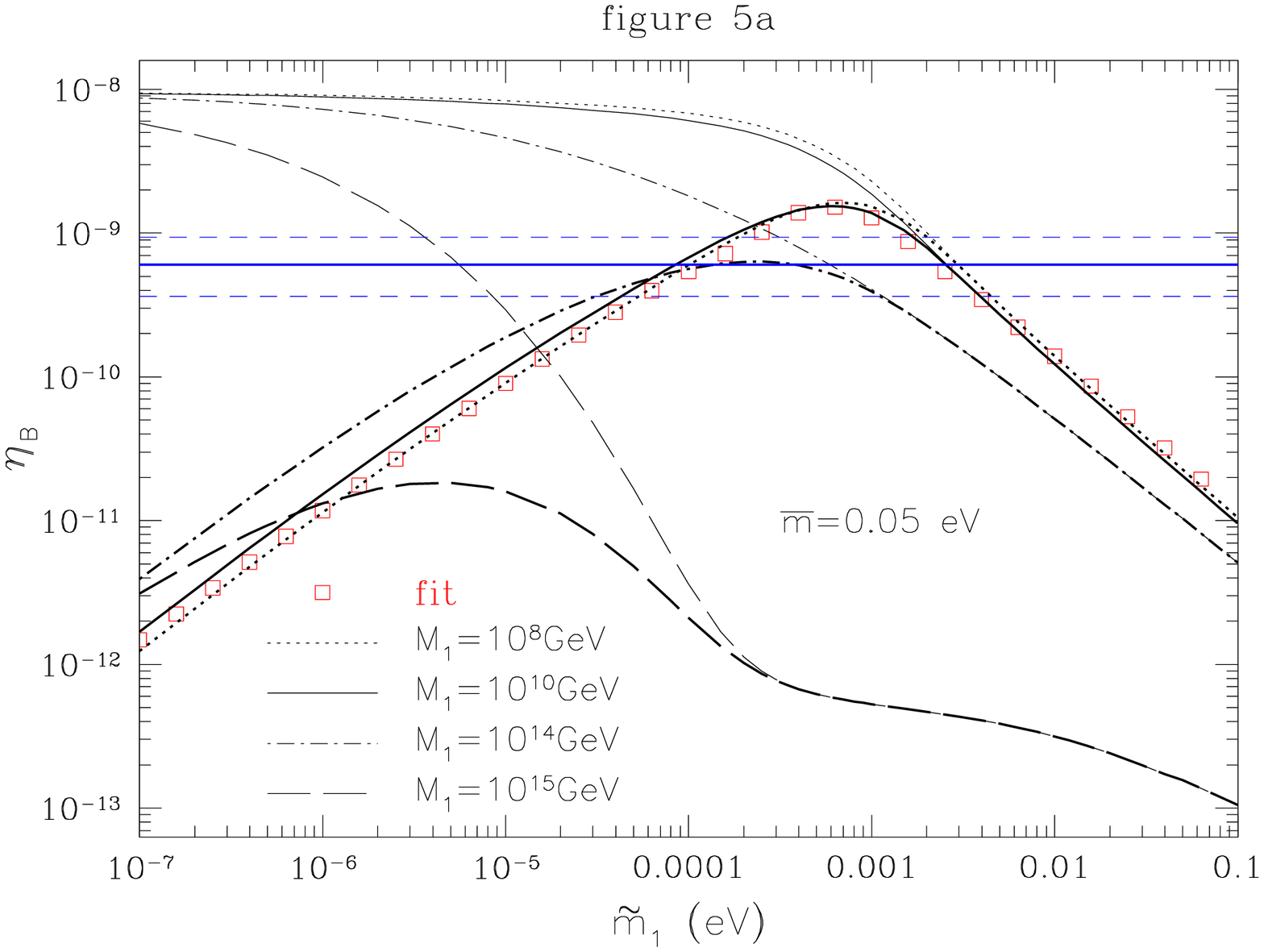,width=112mm}

\psfig{file=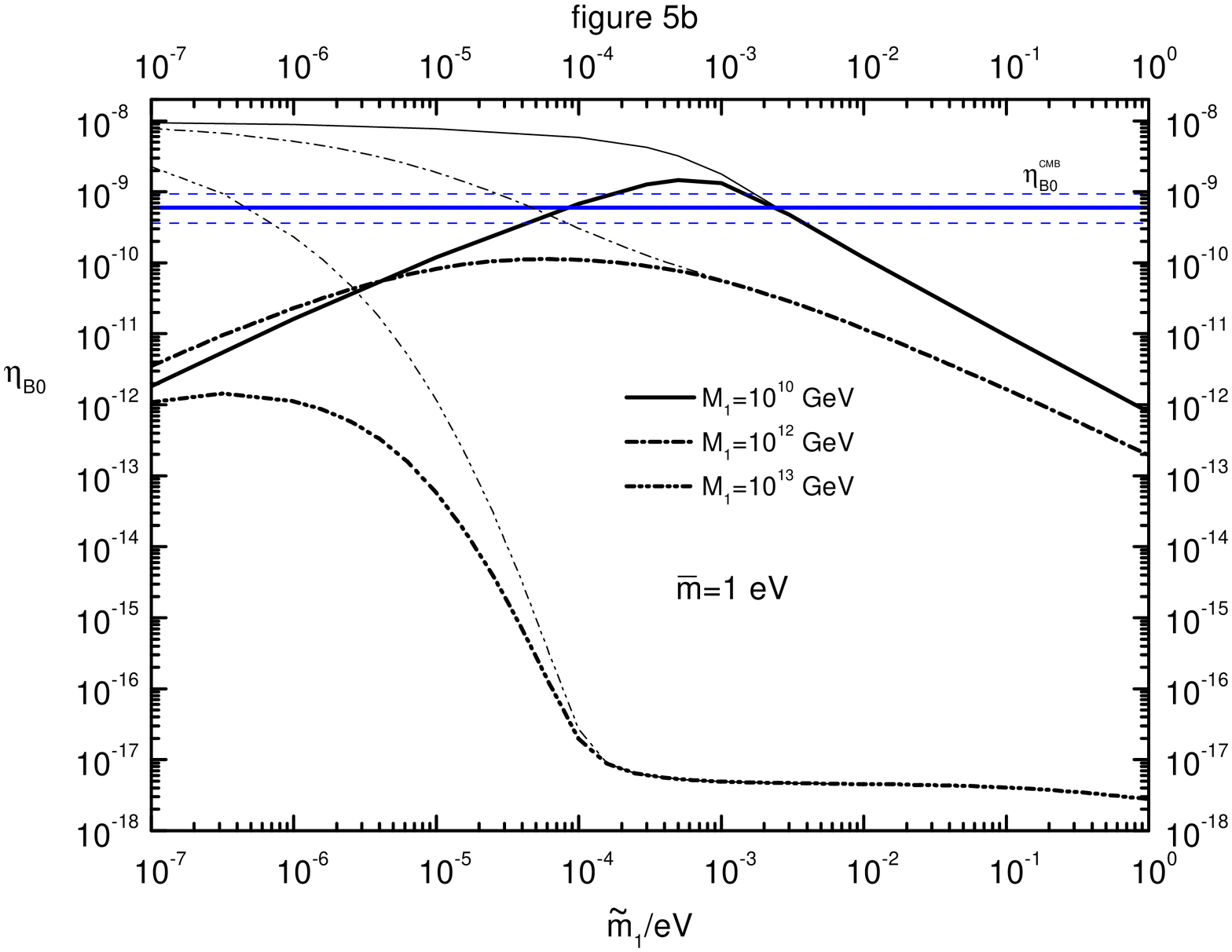,width=14cm}

{\bf Figure 5.} Predicted baryon-to-photon ratio $\h_{B0}$ as a function of
$\mt$, for $\ve_1=-10^{-6}$ and for the indicated values of $M_1$.
The horizontal solid and dashed lines indicate the mean and the upper/lower values
($3\sigma$) of $\h_{B0}^{CMB}$, respectively (see eq.~(\ref{obs})).
(a) Hierarchical neutrino case ($\mb=0.05\ {\rm eV}\simeq \sqrt{\D m^2_{\rm atm}}$).
The squares denote the fit (\ref{fit}).
(b) Quasi-degenerate neutrino case ($m_{\nu_i}\simeq 0.58\ {\rm eV}$).

\newpage
\hspace{11mm}
\psfig{file=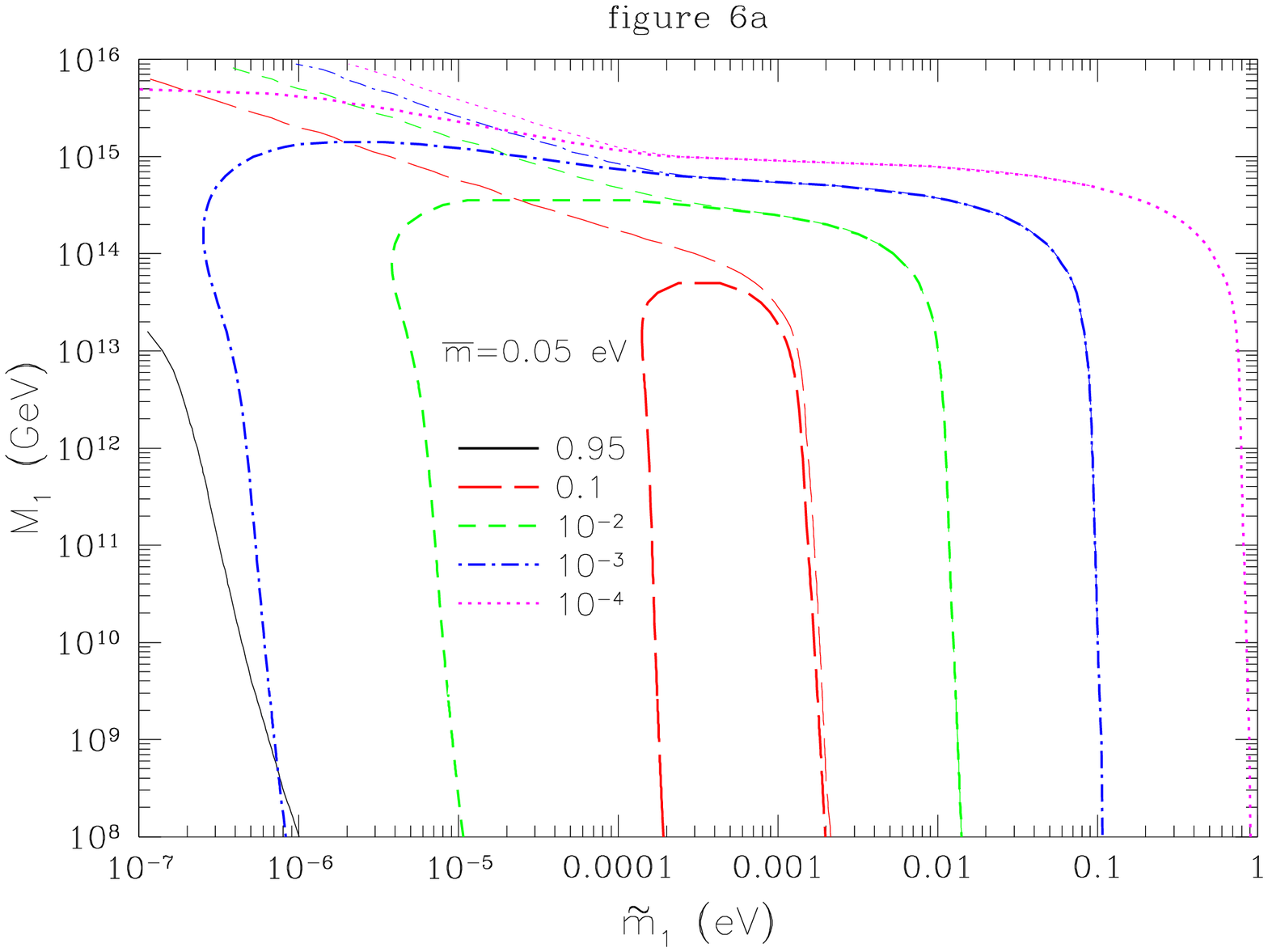,width=112mm}

\psfig{file=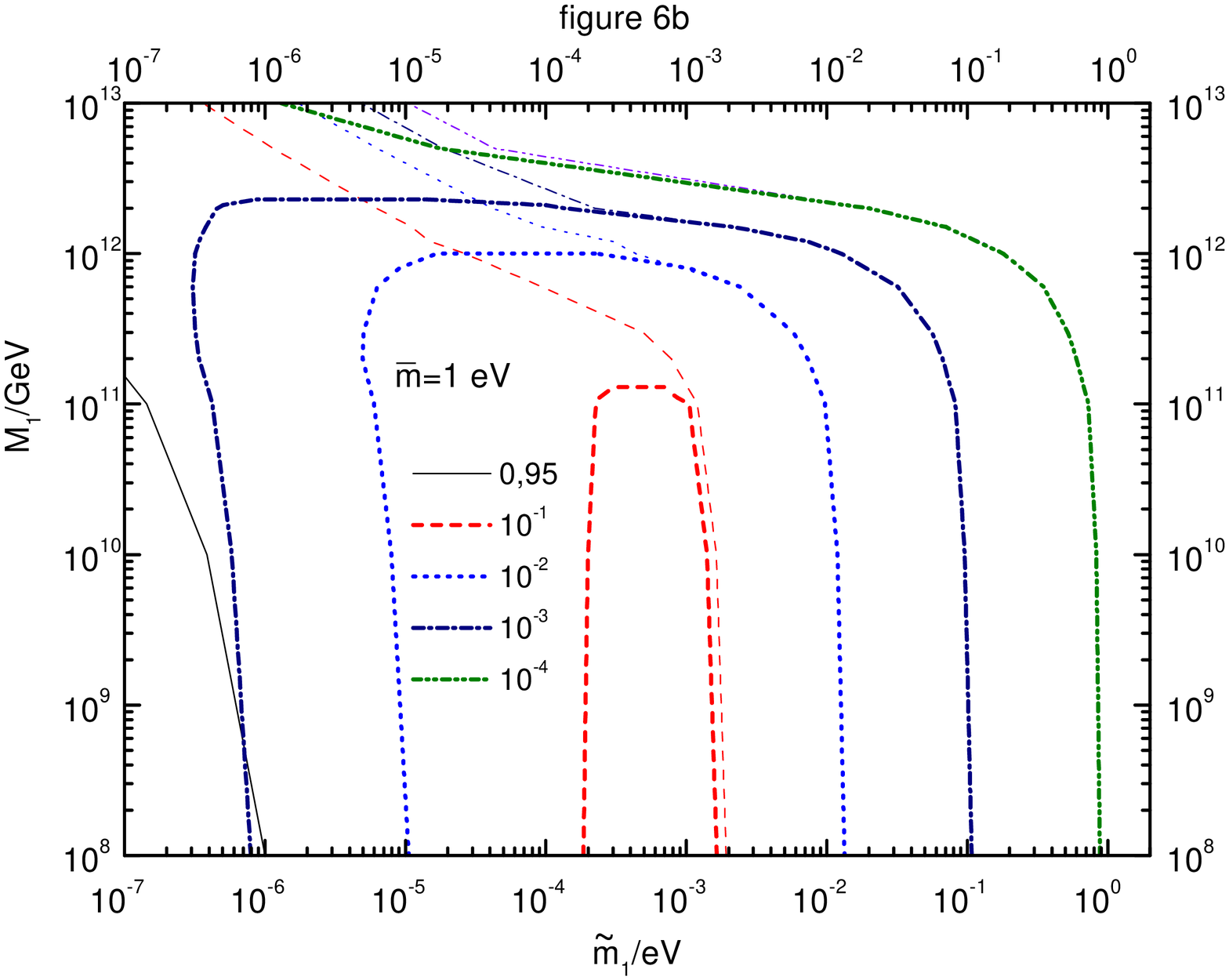,width=14cm}

{\bf Figure 6.} Iso-$\k_0$ curves.
(a) Hierarchical neutrino case ($\mb=0.05\ {\rm eV}\simeq \sqrt{\D m^2_{\rm atm}}$).
(b) Quasi-degenerate neutrino case ($m_{\nu_i}\simeq 0.58\ {\rm eV}$).

\newpage
\hspace{11mm}
\psfig{file=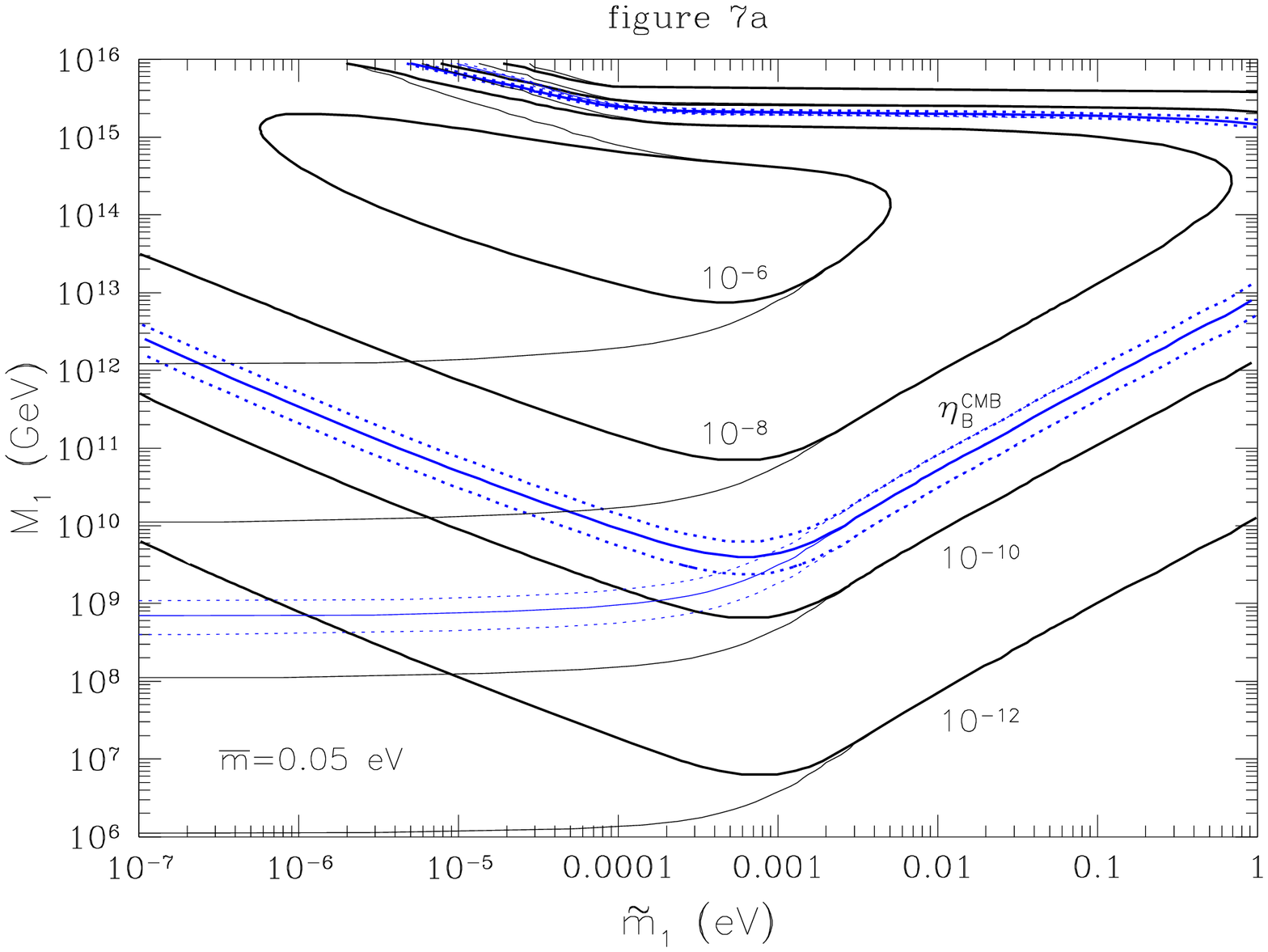,width=112mm}

\psfig{file=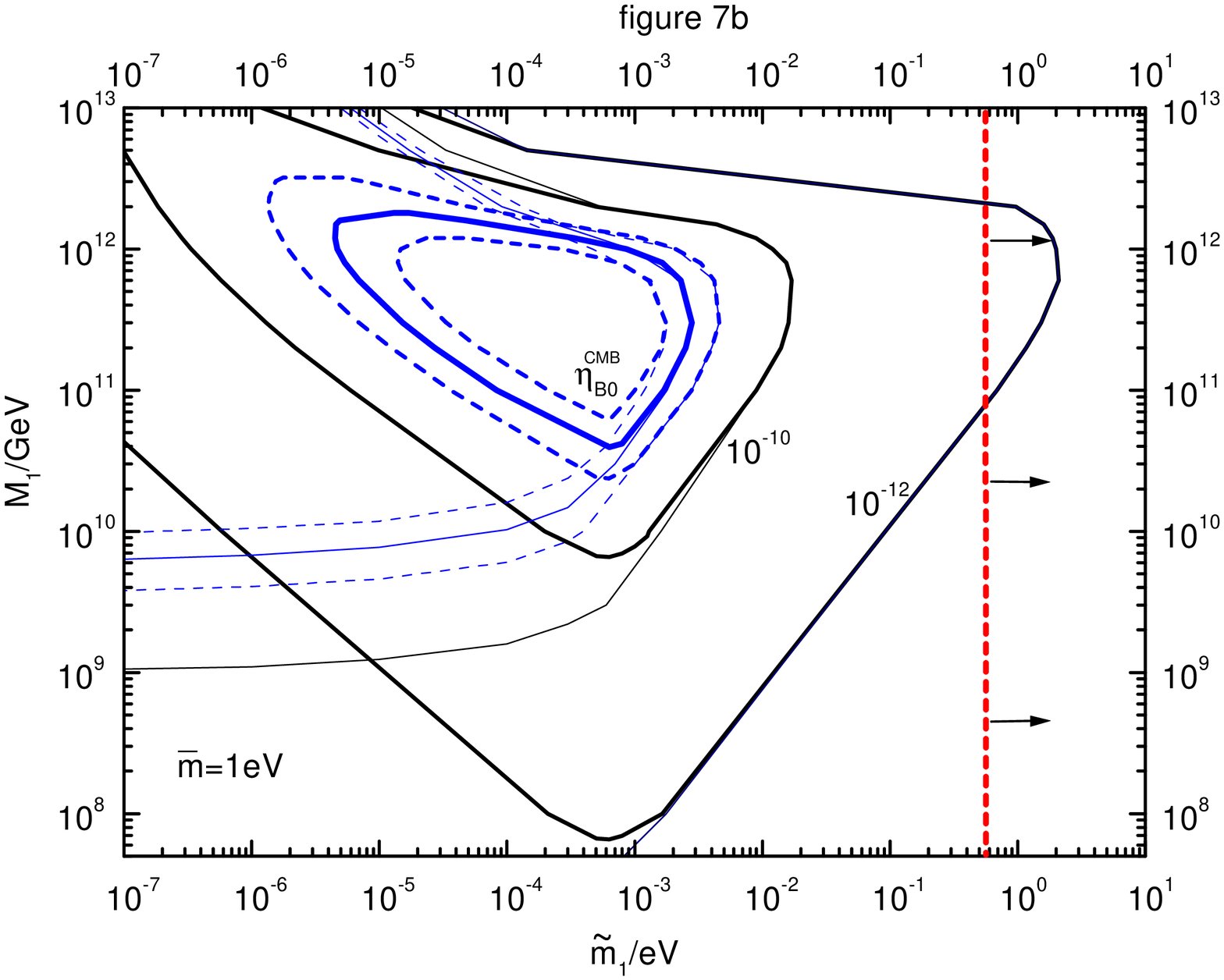,width=14cm}

{\bf figure 7.} Iso-$\h_{B0}^{\rm max}$ curves.
The region within the dashed lines is currently allowed by the CMB constraint
(cf.~\ref{constraint}) for the ($3\s$) lower $\eta_{B0}^{CMB}$ value.
(a) Hierarchical neutrino case ($\mb=0.05\ {\rm eV}\simeq \sqrt{\D m^2_{\rm atm}}$).
(b) Quasi-degenerate neutrino case ($m_{\nu_i}\simeq 0.58\ {\rm eV}$).
The dashed vertical line and the arrows refer to the bound (\ref{mt1lbound}).

\newpage
\centerline{\underline{\Large\bf Erratum}}
%

%\author{W.~Buchm\"uller, P. Di Bari\\
%{\it Deutsches Elektronen-Synchrotron DESY, 22603 Hamburg, Germany}\\[5ex]
%M.~Pl\"umacher\\
%{\it Theoretical Physics, University of Oxford, 1 Keble Road,}\\
%{\it Oxford, OX1 3NP, United Kingdom}
%}

\maketitle

\vspace{5mm}
% \centerline{\date{\today}}

As pointed out by Giudice et al.\ \cite{giudice}, a proper
treatment of the real intermediate state contribution to the
$\Delta L=2$ scattering processes also subtracts the term given in
Eq.~(18). This corrects a result arising from the subtraction
procedure introduced in \cite{early}, which has been extensively
used in the literature. Removing the spurious term (18) reduces
the washout rate $\Gamma_W$ by a factor of about $2/3$. As a
consequence, the numerical factors change in Eqs. (60) and (63):
\begin{displaymath}
  M_1\gtrsim 2.2\,(0.4)\times10^9\,{\rm GeV}
  \left(\frac{2.5\times10^{-3}\,
  {\rm eV}^2}{\Delta m_{\rm atm}^2}\right)^{1/2}\;,
\end{displaymath}
and
\begin{displaymath}
M_1 \simeq 2\times 10^{10}\,{\rm GeV}\, \left(\widetilde{m}_1\over
0.01\,{\rm eV}\right) \, .
\end{displaymath}
A detailed discussion can be found in \cite{annals}.

\end{document}